\PassOptionsToPackage{nameinlink}{cleveref}
\documentclass[a4paper,UKenglish,cleveref,autoref,thm-restate,hideLIPIcs,nolineno]{lipics_no_doi}

\usepackage{latexsym,amsmath,amsthm,amssymb}

\usepackage{pifont}
\usepackage{enumerate}
\usepackage{subcaption}
\usepackage{cprotect}
\usepackage[T1]{fontenc}
\usepackage{bbm}
\usepackage{cite}
\usepackage{mathtools}
\usepackage{amsfonts}
\usepackage{pifont}
\usepackage{url}
\usepackage{multirow}
\usepackage{multicol}
\usepackage{fancybox}
\usepackage{colortbl}
\usepackage{soul}
\usepackage{color}
\usepackage{tikz}
\usepackage{tkz-berge}
\usetikzlibrary{calc} 
\usetikzlibrary{backgrounds}
\usetikzlibrary{arrows.meta}
\usepackage{rotating}
\usepackage{pdflscape}
\usepackage{afterpage}
\usepackage{comment}
\usepackage{etoolbox}
\usepackage{environ}
\usepackage{dirtytalk}
\usepackage[rgb]{xcolor}

\usepackage{cite}

\usepackage{algorithm}
\usepackage{algpseudocode}

\usepackage{todonotes}

\definecolor{mid-green}{rgb}{0.15,0.65,0.15}
\definecolor{dark-green}{rgb}{0.15,0.25,0.15}
\definecolor{dark-red}{rgb}{0.7,0.15,0.15}
\definecolor{dark-blue}{rgb}{0.15,0.15,0.9}
\definecolor{medium-blue}{rgb}{0,0,0.5}
\definecolor{gray}{rgb}{0.5,0.5,0.5}
\definecolor{color-Ig}{rgb}{0.15,0.7,0.15}
\definecolor{darkmagenta}{rgb}{0.30, 0.0, 0.30}

\hypersetup{
  colorlinks, linkcolor={dark-blue},
citecolor={mid-green}, urlcolor={dark-blue}}

\Crefname{remark}{Remark}{Remarks}
\Crefname{lemma}{Lemma}{Lemmas}

\Crefname{invariant}{Invariant}{Invariants}

\Crefname{fvsfvsrule}{Rule}{Rules}

\newtheorem{vcvcrule}{Reduction Rule}

\Crefname{vcvcrule}{Rule}{Rules}

\usetikzlibrary{decorations,arrows,shapes}
\newcommand{\inners}{1.2pt}
\newcommand{\outers}{1pt}

\newcommand{\pdkernelname}{PDE}
\newcommand{\pdkernel}{{\pdkernelname{} kernel}}
\newcommand{\pdkernels}{{\pdkernelname{} kernels}}

\usepackage{complexity}
\newcommand{\Hard}{\text{hard}}
\newclass{\pNP}{paraNP}
\newcommand{\Hness}{hardness}
\newcommand{\NPH}{\NP\text{-}\Hard}
\newcommand{\NPHness}{\NP\text{-}\Hness}

\newclass{\Complete}{complete}
\newclass{\Total}{Total}
\newclass{\Delay}{Delay}
\newclass{\Cness}{completeness}
\newclass{\PPT}{PPT}
\newclass{\IncP}{IncP}
\newclass{\DelayP}{DelayP}
\newclass{\Pclass}{P}

\newfunc{\super}{super}
\newfunc{\tin}{in}
\newfunc{\tout}{out}
\newfunc{\bits}{bits}
\newfunc{\Conf}{conf}

\newfunc{\Sol}{Sol}
\newfunc{\YES}{yes}
\newfunc{\NOi}{no}
\newfunc{\bdd}{bd}
\newfunc{\tw}{tw}
\newfunc{\td}{td}
\newfunc{\dcc}{dcc}
\newfunc{\dc}{dc}
\newfunc{\roott}{root}
\newfunc{\mul}{mul}
\newfunc{\fvs}{fvs}
\newfunc{\dirx}{dir}
\newfunc{\allx}{all}
\newfunc{\largex}{large}
\newfunc{\prop}{propagate}
\newcommand{\propDir}{\prop_\dirx}
\newcommand{\propAll}{\prop_\allx}
\newcommand{\propLarge}{\prop_\largex}
\newclass{\Wc}{W}

\newfunc{\dist}{dist}
\newfunc{\diam}{diam}
\newfunc{\border}{border}
\newfunc{\dgr}{deg}
\newfunc{\degR}{degR}

\newfunc{\Rep}{Rep}
\newfunc{\DPx}{DP}

\newfunc{\leavesx}{leaves}

\newcommand{\pname}[1]{{\sc #1}}

\newfunc{\opt}{opt}
\newfunc{\rmcx}{rmc}

\newfunc{\ins}{ins}
\newfunc{\shift}{shift}
\newfunc{\glue}{glue}
\newfunc{\proj}{proj}
\newfunc{\joinf}{join}
\newfunc{\idx}{index}
\newfunc{\Lift}{Lift}
\newfunc{\ePPT}{ePPT}
\newfunc{\up}{up}
\newfunc{\cbd}{cb}
\newfunc{\down}{down}

\def\A{\mathcal{A}}
\def\C{\mathcal{C}}
\def\D{\mathcal{D}}

\def\R{\mathcal{R}}

\newcommand{\inpFPT}{input-\(\FPT\)}

\newcommand{\bigO}[1]{{\mathcal{O}\!\left(#1\right)}}

\newcommand{\problenum}[3]{
  \begin{flushleft}
    \fbox{
      \begin{minipage}{0.97\linewidth}
        \noindent {\sc #1}\\
        {\bf Instance:} #2\\
        {\bf Enumerate:} #3
      \end{minipage}}
    \medskip
  \end{flushleft}
}

\newenvironment{sproof}[1][\todowarn]
  {\begin{proof}[Proof of safeness and lifting algorithm of~#1]}
  {\end{proof}}
  
\newenvironment{cproof}
  {\begin{proof}[Proof of the claim]}
  {\end{proof}}

\definecolor{goodred}{HTML}{CC6677}
\definecolor{goodblue}{HTML}{332288}
\definecolor{goodyellow}{HTML}{DDCC77}
\definecolor{goodgreen}{HTML}{117733}
\definecolor{goodcyan}{HTML}{88CCEE}
\definecolor{goodwine}{HTML}{882255}
\definecolor{goodteal}{HTML}{44AA99}
\definecolor{goodolive}{HTML}{999933}
\definecolor{goodpurple}{HTML}{AA4499}

\newcommand{\todowarn}{\PackageWarning{todonotes}{something to be done there.}}

\algnewcommand\algorithmicforeach{\textbf{for each}}
\algdef{S}[FOR]{ForEach}[1]{\algorithmicforeach\ #1\ \algorithmicdo}

\newdimen{\algindent}
\algrenewcommand\algorithmicindent{\algindent}%
\algnewcommand\LeftComment[2]{%
\hspace{#1\algindent}$\triangleright$ #2 \hfill%
}

\usetikzlibrary{decorations.pathmorphing}

\title{Enumeration kernels for Vertex Cover and Feedback Vertex Set} 

\author{Marin Bougeret}{LIRMM, Université de Montpellier, Montpellier,
France}{marin.bougeret@lirmm.fr}{https://orcid.org/0000-0002-9910-4656}{}

\author{Guilherme C. M. Gomes}{LIRMM, Université de Montpellier, CNRS, Montpellier,
France\\
Universidade Federal de Minas Gerais, Belo Horizonte, Brazil}{gcm.gomes@dcc.ufmg.br}{https://orcid.org/0000-0002-7487-3475}{Funded by the European Union, project PACKENUM, grant number 101109317. Views and opinions expressed are however those of the author only and do not necessarily reflect those of the European Union. Neither the European Union nor the granting authority can be held responsible for them.}

\author{Vinicius F. dos Santos}{Universidade Federal de Minas Gerais, Belo Horizonte, Brazil}{}{https://orcid.org/0000-0002-4608-4559}{Supported by CNPq (Grants 312069/2021-9, 406036/2021-7, 404479/2023-5, and 308129/2025-3)}

\author{Ignasi Sau}{LIRMM, Université de Montpellier, CNRS, Montpellier,
France}{ignasi.sau@lirmm.fr}{https://orcid.org/0000-0002-8981-9287}{French project ELIT (ANR-20-CE48-0008-01).}

\authorrunning{ M. Bougeret, G. C. M. Gomes, V. F. dos Santos, and I. Sau} 

\Copyright{Marin Bougeret, Guilherme C. M. Gomes, Vinicius F. dos Santos, and Ignasi Sau} 



\category{} 

\relatedversion{} 

\begin{document}

\maketitle

\begin{abstract}
Enumerative kernelization is a recent and promising area sitting at the intersection of parameterized complexity and enumeration algorithms.
Its study began with the paper of Creignou et al. [Theory Comput. Syst., 2017], and development in the area has started to accelerate with the work of Golovach et al. [J. Comput. Syst. Sci., 2022].
The latter introduced polynomial-delay enumeration kernels and applied them in the study of structural parameterizations of the \textsc{Matching Cut} problem and some variants.
Few other results, mostly on \pname{Longest Path} and some generalizations of \textsc{Matching Cut}, have also been developed.
However, little success has been seen in enumeration versions of \textsc{Vertex Cover} and \textsc{Feedback Vertex Set}, some of the most studied problems in  kernelization.
In this paper, we address this shortcoming. 
Our first result is a polynomial-delay enumeration kernel with $2k$ vertices for \textsc{Enum Vertex Cover}, where we wish to list all solutions with at most $k$ vertices.
This is obtained by developing a non-trivial lifting algorithm for the classical crown decomposition reduction rule, and directly improves upon the kernel with $\mathcal{O}(k^2)$ vertices derived from the work of Creignou et al.
Our other result is a polynomial-delay enumeration kernel with $\mathcal{O}(k^3)$ vertices and edges for \textsc{Enum Feedback Vertex Set}; the proof is inspired by some ideas of Thomassé [TALG, 2010], but with a weaker bound on the kernel size due to difficulties in applying the $q$-expansion technique.
\end{abstract}

\section{Introduction}
\label{sec:intro}

Enumeration problems are central objects in both practical and theoretical computer science, with the main example of the former being the branch-and-bound method~\cite{branch_bound} (and its relatives), universally used in integer programming solvers.
Instead of answering a question in either the positive or the negative, in this class of problems the goal is to generate all witnesses, or \emph{solutions}, that satisfy the given problem's constraints, or decide that no solution exists; we denote by $\Sol(x)$ the solution set of an instance $x$ of some fixed problem of interest.
Typically, the number of solutions will be exponential in the input size $n$, and thus \emph{input-sensitive} algorithms, whose running-times are only analyzed with respect to $n$, are not powerful enough tools to capture all nuances intrinsic to enumeration problems.
In particular, \emph{input-polynomial} algorithms, whose running times are of the form $n^\bigO{1}$, are inadequate models of efficient enumeration.
Consequently, enumeration algorithms are usually analyzed in the \emph{output-sensitive} context, i.e., their running times depend on both $n$ and the size of the output.
In this setting, a popular class of efficiently solvable enumeration problems are those that admit \emph{polynomial-delay}~\cite{bagan_delay,fo_query_delay,maximal_ind_set_delay} algorithms, where the running time between consecutive outputs must be bounded by $\poly(n)$.
These algorithms are unfeasible goals in many interesting cases, as enumeration problems for which the existence of a single solution is an $\NPH$ problem trivially do not admit such algorithms unless $\Pclass = \NP$.

Orthogonally to enumeration, the theory of parameterized complexity~\cite{cygan_parameterized,downey_fellows,book_kernels} gained significant traction as a means of coping with \NPHness.
In this framework, an instance $x$ is equipped with a \emph{parameter} of interest $k = \kappa(x)$ and the efficient algorithms, known as \emph{fixed-parameter tractable} (\FPT), are those of running time $f(k) \cdot n^\bigO{1}$, for some computable function $f$.
An equivalent definition of tractability in this framework, and central to our work, is that of \emph{decision kernelization}: a polynomial-time algorithm, or \emph{kernel}, that compresses an instance $(x,k)$ into an equivalent instance $(y,\ell)$  such that $|y|,\ell \leq g(k)$ for some computable function $g$, known as the \emph{size} of the kernel.
Beyond their theoretical interest, kernels are also models for theoretically guaranteed \emph{pre-processing} algorithms, a key feature in real-world applications, such as in integer optimization.
Polynomial kernels, i.e., when $g(k) = \poly(k)$, are of particular interest, as they typically result in significant reductions in input size.
Kernelization theory counts with several robust results and tools~\cite{crown_decomp_fellows,thomasse_fvs,subquad_implicit,rainbow_matching,book_kernels,dynamic_hit_pack,bridgedepth,cross_composition,packing_kernel}, including meta-theorems~\cite{meta_kernel,dp_friendly,bidimensionality,protrusions_sau}, for both upper and lower bounds.

A very successful case study in kernelization is that of \pname{Vertex Cover}, arguably the most fundamental problem in parameterized complexity.
Abu-Khzam et al.~\cite{crown_decomp_fellows} presented the first kernel with $\bigO{k}$ vertices for the problem when parameterized by the solution size $k$, using the then-novel concept of \emph{crown decompositions} to improve on the previously best known kernel with $\bigO{k^2}$ vertices, implied by a result of Buss and Goldsmith~\cite{buss_kernel}.
Currently, the best known kernels for \pname{Vertex Cover} have size $2k$~\cite{nemhauser_crown}, and are based on the half-integral relaxation of Nemhauser and Trotter~\cite{nemhauser_trotter}.
Alongside \pname{Vertex Cover}, the \pname{Feedback Vertex Set} problem has been quite influential in kernelization theory~\cite{fvs_k11,fvs_k3,thomasse_fvs}. 
When parameterized by the solution size $k$, it was first shown to admit a kernel with $\bigO{k^{11}}$ vertices by Burrage et al.~\cite{fvs_k11}; this was later improved to a cubic kernel by Bodlaender and van Dijk~\cite{fvs_k3}.
Shortly afterwards, Thomassé~\cite{thomasse_fvs} presented a kernel with $\bigO{k^2}$ vertices and edges by introducing a generalization of crown decompositions known as \emph{$q$-expansions}, which have since been used in other examples in the literature~\cite{subquad_implicit,enum_long_path,minor_hitting,coc_kumar}.
Thomassé's kernel size has also been shown to be asymptotically optimal by Dell and Marx~\cite{packing_kernel}.

Over the last years, there have been mounting efforts to extend parameterized complexity to the enumeration setting~\cite{meeks_oracle,fernau_param_enum,damaschke_cluster_editing,damaschke_full_kernels,golovach2022refined,enum_long_path,multicut_ipec,oscar_degeneracy,enum_dcut,creignou2017enum}, with \emph{\inpFPT} and \emph{\FPT-delay} having their expected definitions; the work of Creignou et al.~\cite{creignou2017enum} has been particularly influential in this endeavor, and we refer to it for a broader discussion on parameterized enumeration. We will also not further discuss input-sensitive enumeration, focusing on delay-based algorithms.
While parameterized enumeration algorithms have been developed for some decades, it took much longer, however, for the area of  \emph{enumerative kernelization} to emerge, even though kernelization itself did play a significant role in several works~\cite{fernau_param_enum,damaschke_full_kernels,creignou2017enum}.

Only very recently Golovach et al.~\cite{golovach2022refined} defined \emph{polynomial-delay enumeration (PDE) kernels}, which are equivalent to the existence of \FPT-delay algorithms, much like \FPT{} algorithms and kernels are equivalent in the decision world; previous models, such as full kernels~\cite{damaschke_full_kernels} and enum-kernels~\cite{creignou2017enum}, had issues that made them unsatisfactory definitions for enumerative kernelization.
Intuitively, a \pdkernel{} has two parts: the first, $\A_1$, is the \emph{compression algorithm} and has the same constraints as a decision kernel, while the second $\A_2$, known as the \emph{lifting algorithm}, receives the input $(x,k)$, the output $(y,\ell)$ of $\A_1$, and a solution $Y \in \Sol(y)$, and must output, with $\poly(|x| + |y| + k + \ell)$-delay, a non-empty subset $S_Y \subseteq \Sol(x)$; additionally, the $S_Y$'s must partition $\Sol(x)$.
\pdkernels{} have been applied to some problems~\cite{golovach2022refined,enum_dcut,multicut_ipec,enum_long_path}, mainly restricted to variants of the \pname{Matching Cut} problem; \cite{enum_long_path} is the unique exception, as it investigates an enumeration variant of the \pname{Longest Path} problem; we remark that they apply the $q$-expansion technique to obtain their results.
Despite being stated as an enum-kernel, Creignou et al.'s~\cite{creignou2017enum} $\bigO{k^2}$ kernel for \pname{Enum Vertex Cover} is a \pdkernel{}.
Despite these positive results, the most basic techniques of decision kernelization theory seem quite difficult to adapt to the enumeration setting.
Motivated by this phenomenon, we examine the best known kernels for \pname{Vertex Cover} and \pname{Feedback Vertex Set} and seek to adapt their ideas and techniques to their respective enumeration variants.

\subparagraph*{Our contributions.} Our first result is a \pdkernel{} for \pname{Enum Vertex Cover} with $2k$ vertices, and directly improves upon the $\bigO{k^2}$ kernel of Creignou et al.~\cite{creignou2017enum}; the problem is formally defined as follows.

\problenum{Enum Vertex Cover}{A graph $G$ and an integer $k$.}{All vertex covers of $G$ of size at most $k$.}

Our proof is a direct generalization of the kernel with $2k$ vertices for \pname{Vertex Cover}; this, in turn, is based on the celebrated crown decomposition technique. Specifically, we obtain the following theorem.

\begin{restatable}{theorem}{thmvcvcstronglinear}
    \label{thm:vcvcstronglinear}
    \pname{Enum Vertex Cover} admits a \pdkernel{} with at most $2k$ vertices when parameterized by the maximum desired size of a solution $k$. 
\end{restatable}

Intuitively, our kernel works by applying the classic crown reduction that deletes the head $H$ and crown $C$ of the decomposition, reduces $k$ by $|H|$, and returns only the body $B$.
The $2k$ bound is consequence of the Nemhauser-Trotter decomposition theorem~\cite{nemhauser_trotter} and the fact that it either yields a crown decomposition or gives a bound on the size of the input graph.
The complicated part is showing how to design a lifting algorithm; we do so by proving that \pname{Enum Vertex Cover} restricted to \emph{crowned graphs}, i.e., graphs that can be partitioned into $H \cup C$ where $C$ is an independent set and there is an $H$-saturating matching between $H$ and $C$, admits a polynomial-delay algorithm.
Given a solution $S$ of the reduced instance, our lifting algorithm has three phases: (i) brute force how the \emph{slack} $k - (|S| + |N(B \setminus S) \cap H|)$ is distributed in $C$, (ii) compute sets of vertices forced to be added to and forbidden in the solution given the output of (i), and (iii) solve the instance $(G', |V(G')|/2)$ induced by the remaining vertices.
The key observations are that: (a) every configuration tested in (i) leads to a solution of the input, and (b) $G'$ is a crowned graph with a perfect matching.

We then proceed to show a cubic \pdkernel{} for the \pname{Enum Feedback Vertex Set} problem, formally defined as below.

\problenum{Enum Feedback Vertex Set}{A multigraph $G$ and an integer $k$.}{Every feedback vertex set of $G$ of size at most $k$.}

We follow a similar direction to Thomassé's quadratic kernel~\cite{thomasse_fvs}: we first design reduction rules to lower bound the minimum degree of the instance; then, we show that upper bounding the maximum degree is enough to lead to a kernel; and finally, we present reduction rules that yield said upper bound.
The core ingredient of Thomassé's proof -- the computation of a $2$-expansion in an auxiliary graph -- seems to fail in the enumeration setting, and we are unable to generalize it.
Nevertheless, we obtain \cref{thm:fvsfvs_strong_kernel}. Our set of reduction rules is mostly different, which leads to a quadratic instead of a linear bound on the maximum degree.
As our lower bound on the minimum degree is also different, we must prove a slight generalization on the bound used in Thomassé's proof.

\begin{restatable}{theorem}{fvsfvsstrongkernel}
    \label{thm:fvsfvs_strong_kernel}
    There is a \pdkernel{} for \pname{Enum Feedback Vertex Set} parameterized by the size of the solution $k$ with $\bigO{k^3}$ vertices.
\end{restatable}

\subparagraph{Organization.} In \autoref{sec:prelim} we present some basic preliminaries about graphs, parameterized complexity, and enumeration problems. In \cref{sec:vcvc_strong}, we present our linear \pdkernel{} for \pname{Enum Vertex Cover}.
In \cref{sec:fvs}, we present our cubic \pdkernel{} for \pname{Enum Feedback Vertex Set} and discuss some challenges in using rules based on $q$-expansions.
We present concluding remarks and possible research directions in \cref{sec:final}.

\section{Preliminaries}
\label{sec:prelim}

We denote the set $\{1, 2, \dots, n\}$ by $[n]$.
We use standard graph-theoretic notation, and we consider simple undirected graphs without loops or multiple edges; see~\cite{murty} for any undefined terminology.
When the graph is clear from the context, the degree (that is, the number of neighbors) of a vertex $v$ is denoted by  $\deg(v)$, and the number of neighbors of a vertex $v$ in a set $A \subseteq V(G)$ and its neighborhood in it are, respectively, denoted by $\deg_A(v)$ and $N_A(v)$; we also define $N(S) = \bigcup_{v \in S} N(v) \setminus S$.
A \textit{matching} $M$ of a graph $G$ is a subset of edges of $G$ such that no vertex of $G$ is incident to more than one edge in $M$; for simplicity, we define $V(M) = \bigcup_{uv \in M} \{u,v\}$ and refer to it as the set of \textit{$M$-saturated vertices}.
The \textit{subgraph of $G$ induced by $X$} is defined as $G[X] = (X, \{uv \in E(G) \mid u,v \in X\})$.
A \emph{feedback vertex set} $X \subseteq V(G)$ is such that $G \setminus X \coloneq G[V(G) \setminus X]$ is a forest; the \emph{feedback vertex number} of $G$ is the size of a feedback vertex set of minimum size.
Given adjacent $u,v \in V(G)$, a \emph{contraction of $u$ into $v$} is an operation that outputs the graph $G \setminus \{u\}$ with the added edges $\{vw \mid w \in N_G(u) \setminus \{v\}\}$.
An algorithm is said to run with \emph{polynomial delay}~\cite{creignou_hard} if the times before the first output (called the \emph{precalculation period}), after the last output (\emph{the postcalculation period)}, and in between two consecutive outputs are bounded by a polynomial in the input size only.

We refer the reader to~\cite{downey_fellows,cygan_parameterized} for basic background on parameterized complexity and present here only definitions pertaining to parameterized enumeration.

\begin{definition}[Parameterized enumeration problem (Creignou et al.~\cite{creignou2017enum}, Golovach et al.~\cite{golovach2022refined})]
    \label{def:enum_problem}
    A \emph{parameterized enumeration problem} over a finite alphabet $\Sigma$ is a tuple $\Pi = (L, \Sol, \kappa)$, shorthanded by $\Pi_\kappa$, such that:
    \begin{enumerate}[i.]
        \item $L \subseteq \Sigma^\star$ is a decidable language;
        \item $\Sol: \Sigma^\star \mapsto 2^{\Sigma^\star}$ is a computable function such that, for every \emph{instance} $x \in \Sigma^\star$, $\Sol(x)$ is finite and we have $\Sol(x) \neq \emptyset$ if and only if $x \in L$;
        \item $\kappa: \Sigma^\star \mapsto \mathbb{N}$ is the \emph{parameterization}.
    \end{enumerate}
    An instance $x$ of $\Pi_\kappa$ is a \emph{\NOi-instance} if $\Sol(x) = \emptyset$, and is a \emph{\YES-instance} otherwise.
\end{definition}

\begin{definition}[Polynomial-delay enumeration kernel]
    \label{def:enum_kernels}
    Let $\Pi_\kappa$ be a parameterized enumeration problem.
    A \emph{polynomial-delay enumeration kernel}, \emph{\pdkernel{}} for short, is a pair of algorithms $\A_1, \A_2$ such that, given an instance $x$ of $\Pi_\kappa$ and $\kappa(x)$:
    \begin{itemize}
        \item The \emph{compression algorithm} $\A_1$ runs in $\poly(|x| + \kappa(x))$-time and outputs an instance $y$ of $\Pi_\kappa$  satisfying $\kappa(y), |y| \leq f(\kappa(x))$, with $f$ being a computable function, and with $\Sol(y) \neq \emptyset$ if and only if $\Sol(x) \neq \emptyset$;
        \item The \emph{lifting algorithm} $\A_2$ receives $x$, $y = \A_1(x,\kappa(x))$, $\kappa(x)$, $\kappa(y)$, some $Y \in \Sol(y)$, and outputs, with polynomial delay on its input, a non-empty $S_Y \subseteq \Sol(x)$.
        Moreover, $\{S_Y \mid Y \in \Sol(y)\}$ is a partition of $\Sol(x)$.
    \end{itemize}
\end{definition}



A \emph{graph subset problem} is a problem with a graph as input and a solution being a subset of its vertices. We say that a \pdkernel{} for a graph subset problem is \emph{extension-only} if, given a solution $Y \in \Sol(y)$, the lifting algorithm works by only adding some vertices in $x \setminus y$ to $Y$.
Intuitively, this constraint implies that a solution of the input instance $x$ cannot be generated by two different solutions of the compressed instance $y$, so it only remains to prove that every solution of $x$ is generated from some solution of $y$ and every solution of $y$ leads to some solution of $x$.
Both kernels developed in this work are extension-only.
We remark that the kernels for \pname{Enum Matching Cut} and its variants developed by Golovach et al.~\cite{golovach2022refined} are not extension only, as their lifting algorithms remove some vertices of the compressed instance's solutions.
\section{A kernel with $2k$ vertices for \pname{Enum Vertex Cover}}
\label{sec:vcvc_strong}

In this section we \cref{thm:vcvcstronglinear}. Our main tools for this result are crowned graphs and crown decompositions, given in \cref{def:crowned_graph}.

\begin{definition}
  \label{def:crowned_graph}
  A graph is a \emph{crowned graph} if its vertex set can be partitioned into $H \cup C$, such that $C$ is an independent set, and such that there is a matching $M$ between $H$ and $C$
  where $|M|=|H|$. The set $C$ is called the \emph{crown}, and $H$ is called the \emph{head}.
  Given a graph $G$, a \emph{crown decomposition}~\cite{cygan_parameterized} of width $t$ is a partition $(C, H, B)$ of $V(G)$ such that $G[H \cup C]$ is
  a crowned graph (with head $H$), $N_G(C)=H$, and $|H| = t$. Set $B$ is called the \emph{body}. 
\end{definition}

To that end, we first reduce the task of finding a \pdkernel{} to an enumeration problem \pname{Enum Crown}, and then reduce \pname{Enum Crown} to
an even more structured enumeration problem \pname{Enum Small Crown} for which we provide a polynomial-delay enumeration algorithm.
Let us first define our problem.

\problenum{Enum Crown}{A crowned graph $G = (H \cup C, E)$ and an integer $x$  in $[0,|C|]$.}{All vertex covers of $G$ of size exactly $|H|+x$.}

The problem \pname{Enum Small Crown} corresponds to the special case of \pname{Enum Crown} where $|C|=|H|$ and $x=0$.
Given an input $(G,x)$ of \pname{Enum Crown}, we denote by $\Sol(G,x)$ the set of solutions.
In the same way, we define $\Sol(G)$ for \pname{Enum Small Crown}.

\begin{remark}\label{obs:enumCrown}
    Every instance of \pname{Enum Crown} has a solution that takes the head and any $x$ vertices in the crown.
    If $M$ is an $H$-saturating matching between $H$ and $C$, then every solution $S$ of \pname{Enum Small Crown} and every $e \in M$ satisfy $|S \cap V(e)|=1$, where $V(e)$ denotes the endpoints of $e$.
\end{remark}

Our kernelization algorithm consists in the exhaustive application of \cref{vcvcrule:deg_0,vcvcrule:crown_reduction}, the latter of which can be applied in polynomial time due to \cref{lemma:crown}.

\begin{vcvcrule}
    \label{vcvcrule:deg_0}
    Let $(G, k)$ be the input instance of \pname{Enum Vertex Cover}. If $v$ is an isolated vertex of $G$, remove $v$ from $G$ and set $k \gets k$.
\end{vcvcrule}

\begin{sproof}[\cref{vcvcrule:deg_0}]
    Let $G' = G \setminus \{v\}$.
    The forward direction is immediate; $S \in \Sol(G,k)$ if and only if $S \setminus \{v\}\in \Sol(G', k)$.
    For the converse direction, it is also immediate that $S' \in \Sol(G',k)$ if and only if $S' \in \Sol(G, k)$; however, observe that $(S' \cup \{v\}) \in \Sol(G, k)$ if and only if $|S'| < k$ as $v$ is isolated, and we must also output $(S' \cup \{v\})$ if possible.
\end{sproof}

As pointed out by Chlebík and Chlebíkova~\cite{nemhauser_crown}, the Nemhauser-Trotter~\cite{nemhauser_trotter} decomposition based on the half-integral relaxation of \pname{Vertex Cover} is  a crown decomposition.

\begin{lemma}[\!\!\cite{nemhauser_trotter} and \cite{nemhauser_crown}]
    \label{lemma:crown}
    Let $G$ be a graph without isolated vertices and with at least $2k+1$ vertices. Then, there is a polynomial-time algorithm that either:
    \begin{itemize}
        \item decides that no half-integral vertex cover of weight at most $k$ exists; or
        \item finds a crown decomposition of $G$ of width at most $k$.
    \end{itemize}
\end{lemma}

\begin{vcvcrule}[Crown reduction]
    \label{vcvcrule:crown_reduction}
    Let $(G,k)$ be an instance of \pname{Enum Vertex Cover} with $(C,H,B)$ a crown decomposition of $G$ of width at most $k$.
    Set $G' \gets G \setminus (H \cup C)$, $k' \gets k - |H|$, and return $(G', k')$.
\end{vcvcrule}

\begin{restatable}{lemma}{lemmacrownimpliesstrong}{}
    \label{lemma:crown_implies_strong}
    If \pname{Enum Crown} has a polynomial-delay enumeration algorithm $\A$, then \cref{vcvcrule:crown_reduction} is a reduction rule with an extension-only lifting algorithm.
\end{restatable}

\begin{proof}
    Let us show how to construct the lifting algorithm $\R_b$ using $\A$.
    Given a solution $S'$ of the reduced instance $(G',k')$, let $x=k'-|S'|$ (where $x \geq 0$) be the \emph{slack} of the solution $S'$, and $F = N(B \setminus S') \cap H$ be the set of \emph{forced} vertices in $H$, i.e. vertices that are adjacent to some vertex of $B$ not picked in $S'$. Let $H'=H \setminus F$.
    For any $\ell \in [0,\min(|C|,x)]$, define $x_\ell=|H'|+\ell$, and observe that $(G[H' \cup C], x_\ell)$ is a valid instance of \pname{Enum Crown} as $x_\ell \le |C|$.
    For any $\ell \in [0,\min(|C|,x)]$, $\R_b$ outputs all solutions of the form $S' \cup F \cup S_\ell$, where $S_\ell \in \Sol(G[H' \cup C],x_\ell)$ are enumerated by $\A$.
    This completes the description of $\R_b$, so let us now prove that it adheres to the requirements of \cref{def:enum_kernels}.  
    The first item is satisfied as it is well known that the crown reduction is safe for the decision version of the problem, in the sense that $(G,k)$ is a \YES-instance if and only if $(G',k')$ is.
    Let us now check the second item, more precisely:
    \begin{itemize}
        \item For every $S' \in \Sol(G',k')$, the set $\R_b(G, G', k, k', S')$ is non-empty, and can be enumerated with polynomial-delay; and
        \item $\{\R_b(G, G', k, k', S') \mid S' \in \Sol(G',k')\}$ is a partition of $\Sol(G,k)$.
    \end{itemize}
    Let us start with the first property.
    Notice that there exists at least one $\ell \in [0,\min(|C|,x)]$, and as for any such $\ell$, tuple $((H',C),x_\ell)$ is an instance of \pname{Enum Crown} (which always
    has a solution by \cref{obs:enumCrown}), we get that $\R_b(G, G', k, k', S') \neq \emptyset$.
    Moreover, $\R_b(G, G', k, k', S')$ runs with polynomial-delay, as for each $\ell$, $\A$ enumerates
    $\Sol(G[H' \cup C], x_\ell)$ with polynomial-delay, and there is no additional verification required in $\R_b$ to avoid duplicate solutions of the form $S' \cup F \cup S_\ell$ generated by different values of $\ell$, as for any $\ell$ these solutions have size exactly $|S'|+|F|+x_\ell$.
    
    Let us now consider the second property.
    Observe first that for any $S' \in \Sol(G',k')$, any $S \in \R_b(G, G', k, k', S')$ is indeed a solution of $(G,k)$ as
    $V(G)$ can be partitioned into $V_1=V(G') \cup F$ and $V_2=H' \cup C$, and $S=S' \cup F \cup S_\ell$ where $S' \cup F$ is a vertex cover of $G[V_1]$,
    $S_\ell$ is a vertex cover of $G[V_2]$, and there is no edge between $V_1 \setminus S$ and $V_2$.
    Let us now prove that $\Sol(G,k) \subseteq \bigcup_{S' \in \Sol(G',k')}\R_b(G, G', k, k', S')$.
    Let $S \in \Sol(G,k)$ and $S' = S \setminus (H \cup C)$. As there is a matching of size $|H|$ between the head and the crown, $|S \cap (H \cup C)| \ge |H|$ and thus $S' \in \Sol(G',k')$.
    Now, let us prove that $S \in \R_b(G, G', k, k', S')$.
    Observe first that as $S$ necessarily contains $F$, $S$ is partitioned into $S=S' \cup F \cup (S \cap (H' \cup C))$, where 
    $S \cap (H' \cup C)$ is a vertex cover of $G[H' \cup C]$.
    Let $\ell_0$ be such that $|S \cap (H' \cup C)|=|H'|+\ell_0$. It remains to prove that $\ell_0 \in [0,\min(|C|,x)]$,
    as this implies that $(S \cap (H' \cup C)) \in \Sol(G[H' \cup C],x_{\ell_0})$, and thus that $S \in \R_b(G, G', k, k', S')$.
    First, we have $\ell_0 \geq 0$ as $(C,H',\emptyset)$ remains a crown decomposition, and thus there is a matching saturating $H'$ in $G[H' \cup C]$.
    Let $x = k'-|S'|$ where $x \geq 0$.
    Moreover, $|S|=|S'|+|F|+ |(S \cap (H' \cup C))|=k'-x+|F|+|H'|+\ell_0=k-x+\ell_0 \le k$, implying that $\ell_0 \le x$. As $\ell_0 \le |C|$ by definition, we get the claimed inequality.
    Finally, observe that $S \cap V(G') = S'$, so it follows that $\R_b$ is an extension-only lifting algorithm, which implies that, for any two different solutions $S'_1,S'_2$ of $\Sol(G',k')$, the sets $\R_b(G, G', k, k', S'_1)$, and $\R_b(G, G', k, k', S'_2)$ are disjoint.
\end{proof}

With this conditional result, we are now able to state a conditional kernelization lemma.

\begin{lemma}\label{lemma:crownImpliesKernel}
    If \pname{Enum Crown} has a polynomial-delay enumeration algorithm, then \pname{Enum Vertex Cover} admits a  \pdkernel{} (with extension-only lifting algorithm) with at most $2k$ vertices.
\end{lemma}

\begin{proof}
    Let $(G,k)$ be an input of \pname{Enum Vertex Cover}.
    We first apply \cref{vcvcrule:deg_0} exhaustively, obtaining $(G_0, k)$.
    If $|V(G_0)| \leq 2k$, then we are done and we return $(G_0,k)$.
    By \cref{lemma:crown}, we can efficiently detect if no fractional vertex cover of weight at most $k$ exists; in this case, we return a trivial \NOi-instance.
    Otherwise, we find a crown decomposition $(C,H,B)$ of $G_0$ using \cref{lemma:crown} and we apply \cref{vcvcrule:crown_reduction}, whose output is $(G_1, k_1)$.
    We repeat this process exhaustively, either returning a \NOi-instance or $(G', k')$.
    This concludes our compression algorithm.
    As we have an extension-only lifting algorithms, they are enough to describe a extension-only \pdkernel; intuitively, it suffices to apply the lifting algorithms of the reduction rules recursively in the reverse order of application order of the rules. More formally, upon receiving a solution $S$, the lifting algorithm of the $j$-th applied reduction rule generates the first solution, say $T_1$, and immediately invokes the lifting algorithm of the $(j-1)$-th applied rule. 
\end{proof}

The remainder of this section is dedicated to proving that \pname{Enum Crown} admits a polynomial-delay enumeration algorithm.
In what follows, given a matching $M$, and a subset $T \subseteq V(M)$ such that for any $e \in M$, $|V(e) \cap T| \leq 1$, we define $M_V(T) \coloneq \{v' \mid vv' \in M \mbox{ and } v \in T\}$ and call it the \emph{image} of $T$ by $M$.
We need the following procedure and its associated technical lemma to obtain our lifting algorithm.
Intuitively, in a small crowned graph $G$, exactly one endpoint of each edge of an $H$-saturating matching $M$ can be in a vertex cover; $\propDir(G, M, X_0)$ decides which vertices must be included and which must be excluded from a solution if we fix $X_0$ to be part of the it.

\begin{figure}
    \centering
    
    \begin{tikzpicture}[scale=0.8]
        \GraphInit[unit=3,vstyle=Normal]
        \SetVertexNormal[Shape=circle, FillColor=black, MinSize=2pt]
        \tikzset{VertexStyle/.append style = {inner sep = \inners, outer sep = \outers}}
        \SetVertexNoLabel
        
        \begin{scope}
            \draw[rounded corners] (0, -0.5) rectangle (7, 0.5);
            
            \node at (-0.75, 0) {$A=H$};
            \foreach \i in {1,...,7} {
                \pgfmathsetmacro{\x}{\i-0.5}
                \Vertex[x=\x, y=0]{h\i}
            }
        \end{scope}

        \begin{scope}[on background layer]
            \fill[goodcyan, rounded corners] ($(h1)-(0.3, 0.3)$) rectangle ($(h2)+(0.3, 0.3)$);
            \node at ($(h1)+(0.5, -0.85)$) {$X_0$};
            
            \fill[goodteal, rounded corners] ($(h3)-(0.3, 0.3)$) rectangle ($(h4)+(0.3, 0.3)$);
            \node at ($(h3)+(0.5, -0.85)$) {$D_2$};

            \draw[thick] ($(h1)-(0,0.65)$) -- ($(h1)-(0,1.15)$) -- ($(h4)-(0,1.15)$) -- ($(h4)-(0,0.65)$);
            \node at ($(h3)-(0.5,1.45)$) {$F$};
        \end{scope}
        
        \begin{scope}[yshift=3cm]
            \draw[dashed, rounded corners] (0, -0.5) rectangle (7, 0.5);
            
            \node at (-0.75, 0) {$B=C$};
            \foreach \i in {1,...,7} {
                \pgfmathsetmacro{\x}{\i-0.5}
                \Vertex[x=\x, y=0]{c\i}
                \Edge[style={-Latex}](h\i)(c\i)
            }
        \end{scope}

        \begin{scope}[on background layer]
            \fill[goodred, rounded corners] ($(c1)-(0.3, 0.3)$) rectangle ($(c2)+(0.3, 0.3)$);
            \node at ($(c1)+(0.5, 0.85)$) {$D_1$};
            
            \fill[goodpurple, rounded corners] ($(c3)-(0.3, 0.3)$) rectangle ($(c4)+(0.3, 0.3)$);
            \node at ($(c3)+(0.5, 0.85)$) {$D_3$};
            
            \draw[thick] ($(c1)+(0,0.65)$) -- ($(c1)+(0,1.15)$) -- ($(c4)+(0,1.15)$) -- ($(c4)+(0,0.65)$);
            \node at ($(c2)+(0.5,1.45)$) {$\bar{F}$};
        \end{scope}

        \begin{scope}[on background layer]
            \fill[goodyellow, rounded corners] ($(h5)-(0.3, 0.3)$) rectangle ($(c7)+(0.3, 0.3)$);
            \node at ($(h7)+(1, 1.5)$) {$G'$};
        \end{scope}
        
        \Edge[style={-latex}](c1)(h2)
        \Edge[style={-latex}](c1)(h3)
        \Edge[style={-latex}](c2)(h4)
        \Edge[style={-latex}](c4)(h1)
    
        \Edge[style={-latex}](c6)(h5)
        \Edge[style={-latex}](c7)(h6)
    \end{tikzpicture}
    \caption{Example of an application of $\propDir(G,  M, X_0)$ that stopped at distance three from $X_0$, with the graph depicted corresponding to $\vec{G}$.
    Edges of $M$ correspond to the vertical arcs, while sloped arcs correspond to all other arcs of $\vec{G}$.
    The light-blue-shaded vertices correspond to $X_0$, the red-shaded to $D_1$ (the vertices at distance one from $X_0$), the teal-shaded to $D_2$, the purple-shaded to $D_3$, and the yellow-shaded vertices correspond to the graph $G' = G[(H \setminus F) \cup (C \setminus \bar{F})]$}.
    \label{fig:propagateAux}
\end{figure}
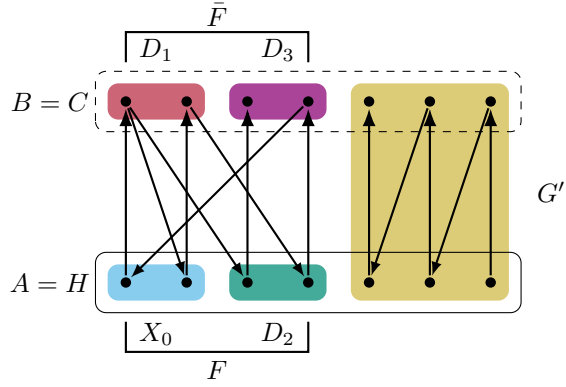

\begin{definition}[$\propDir(G, M, X_0)$]
    \label{def:propaux}
    Given a small crowned graph $G = (H \cup C, E)$ (meaning with $|H|=|C|$) and $X_0 \subseteq H$, the procedure $\propDir(G, M, X_0)$ outputs two sets $(F,\bar{F})$ as follows (see \cref{fig:propagateAux}).
    Let $\vec{G} = (A \cup B, \vec{E})$ be an oriented bipartite graph with $A=H$, $B =C$, where:
    \begin{itemize}
        \item for any edge $e$ between $A$ and $B$ with $e \in M$, we add its oriented version from $A$ to $B$.
        \item for any edge $e$ between $A$ and $B$ with $e \notin M$, we add its oriented version from $B$ to $A$.
    \end{itemize}
    Let $Z \subseteq A \cup B$ be the vertices that can be reached from $X_0$ in $\vec{G}$, including $X_0$ itself.
    Then, $\propDir(G, M, X_0)$ outputs $(F,\bar{F})$ where $F =  Z \cap A$ and $\bar{F}=Z \cap B$. 
\end{definition}


\begin{lemma}\label{lemma:propagateAux}
    Let $G$ be a small crowned graph, $M$ a perfect matching of $G$ between $H$ and $C$, $X_0 \subseteq H$, and $(F,\bar{F})=\propDir(G, M, X_0)$. Then:
    \begin{enumerate}
        \item \label{aux:prop1} $\bar{F}=M_V(F)$ (where $F \subseteq H$ and $\bar{F} \subseteq C$) and for any solution $S \in \Sol(G)$ such that $X_0 \subseteq S$, $S$ contains $F$ and avoids $\bar{F}$.
        \item \label{aux:prop2} For any solution $S \in \Sol(G)$ such that $X_0 \subseteq S$, if we define $H'= H \setminus F$ and $C'= C \setminus \bar{F})$,then $G' = G[H' \cup C']$ is a small crowned graph and $S \cap (C' \cup H') \in \Sol(G')$.
        \item \label{aux:prop3} For any solution $S' \in \Sol(G')$, $F \cup S'$ is a solution of $G$ (that contains $X_0$).
    \end{enumerate}  
\end{lemma}

\begin{proof}
    Let $\vec{G} = (A \cup B, \vec{E})$ be the oriented bipartite graph defined in \cref{def:propaux}.
    
    Proof of \cref{aux:prop1}.
    Let $S \in \Sol(G)$ such that $X_0 \subseteq S$.
    For any $i \ge 0$, let $D_i$ be the vertices of $(A,B)$ at distance $i$ from $X_0$ in $\vec{G}$ (e.g. $D_0 = X_0$, $D_1=M_V(X_0)$, see \cref{fig:propagateAux}).
    Let us show by induction that for any $i$, if $i$ is even, then $D_i \subseteq A \cap S$, and if $i$ is odd then $D_i \subseteq B \setminus S$,
    and $D_{i}=M_V(D_{i-1})$.
    Observe that the statement holds for $i=0$.
    Let us now assume that it holds for $i-1$ and consider $i$.
    If $i$ is odd, observe first that $D_i=M_V(D_{i-1})$ by the definition of $\vec{G}$, and, by \cref{obs:enumCrown}, for any edge $v,v' \in M$ with $v \in D_{i-1}$ and $v' \in D_i$,
    $v' \notin S$.
    If $i$ is even, as $D_{i-1} \cap S = \emptyset$ and $S$ is a vertex cover of $G$, then $D_i \subseteq S$, and, as $\vec{G}$ is bipartite, we also get that $D_i \subseteq A$.
    Now, as $F = \bigcup_{i \equiv 0 \mod 2}D_i$ and $\bar{F} = \bigcup_{i \equiv 1 \mod 2}D_i$, we get $\bar{F}=M_V(F)$, and \cref{aux:prop1} follows.
    
    Proof of \cref{aux:prop2}.
    Observe first that $G'$ is indeed a crowned graph as $\bar{F}=M_V(F)$.
    Now, as $S \in \Sol(G)$ we get $|S|=|H|$; by \cref{obs:enumCrown} we get $|S \cap (F \cup \bar{F})|=|F|$, leading to $|S \cap (C' \cup H')|=|H'|$.
    
    Proof of \cref{aux:prop3}. Let $S' \in \Sol(G')$, and let us consider $S=F \cup S'$.
    Observe that $H \cup C$ is partitioned into $F \cup \bar{F}$ and $H' \cup C'$, that $F$ is a vertex cover of $G[F \cup \bar{F}]$ (as $\bar{F} = M_V(F)$),
    and $S'$ is a vertex cover of $G[H' \cup C']$. By definition of $F$ and $\bar{F}$, there is no edge between $\bar{F}$ and $H' \cup C'$, implying that
    $S$ is a vertex cover of $G[H \cup C]$. 
    Since $|S|=|H|$, we get that $S$ is a solution of $G$ (that contains $X_0$ by definition of $F$).
\end{proof}

\begin{lemma}\label{lemma:SmallCrownBigCrown}
  If \pname{Enum Small Crown} admits a polynomial-delay algorithm, then \pname{Enum Crown} also admits such an algorithm.
\end{lemma}
\begin{proof}
    Let us define a polynomial-delay enumeration algorithm $\A$ for \pname{Enum Crown}.
    Let $(G, x)$ be an instance of \pname{Enum Crown} with $V(G) = H \cup C$, and let $M$ be a matching from $H$ to $C$ that saturates $H$.
    Given $T \subseteq V(G)$, we define $M(T)=\{e \in M \mid V(e) \subseteq T\}$, i.e., the set of edges of $M$ entirely in $T$.
    For any $S \in \Sol(G,x)$, the \emph{signature of $S$} is composed of two sets, $C_1(S) = V(M(S)) \cap C$ and $C_2(S) = (C \setminus V(M)) \cap S)$.
    Observe that
    \begin{itemize}
        \item $|C_1(S)|=|M(S)|$, where $|C_1(S)| \le x$ (as $|C_1(S)|+|H| \le |S| = |H|+x$);
        \item $x-(|C|-|M|) \le |C_1(S)|$ (as if $x-(|C|-|M|) > |C_1(S)|$ then as $C_2(S) \subseteq (C \setminus V(M))$, we would have $|S|=|H|+|C_1(S)|+|C_2(S)| < |H|+x$), and $|C_2(S)|=x-|C_1(S)|$;
        \item By defining $\bar{X_1}(S)=(C \setminus V(M)) \setminus C_2(S)$, we have $S \cap \bar{X_1}(S) = \emptyset$.
    \end{itemize}
    
    
    \begin{figure}[!htb]
        \centering
        
        \begin{tikzpicture}[scale=0.8]
            \GraphInit[unit=3,vstyle=Normal]
            \SetVertexNormal[Shape=circle, FillColor=black, MinSize=2pt]
            \tikzset{VertexStyle/.append style = {inner sep = \inners, outer sep = \outers}}
            \SetVertexNoLabel

            \begin{scope}[yshift=3cm]
                \draw[dashed, rounded corners] (-6, -0.5) rectangle (8, 0.5);
                
                \node at (-6.5, 0) {$C$};
                \foreach \i in {1,...,14} {
                    \pgfmathsetmacro{\x}{\i-6.5}
                    \Vertex[x=\x, y=0]{c\i}
                }
            \end{scope}
    
            \begin{scope}[on background layer]
                \fill[goodcyan, rounded corners] ($(c13)-(0.3, 0.3)$) rectangle ($(c14)+(0.3, 0.3)$);
                \node at ($(c13)+(0.5, 0.85)$) {$C_1$};

                \fill[goodpurple, rounded corners] ($(c7)-(0.3, 0.3)$) rectangle ($(c10)+(0.3, 0.3)$);
                \node at ($(c8)+(0.5, 0.85)$) {$\bar{F}$};
                
                \fill[goodred, rounded corners] ($(c4)-(0.3, 0.3)$) rectangle ($(c6)+(0.3, 0.3)$);
                \node at ($(c5)+(0, 0.85)$) {$\bar{X}_1$};
                
                \fill[goodcyan, rounded corners] ($(c1)-(0.3, 0.3)$) rectangle ($(c3)+(0.3, 0.3)$);
                \node at ($(c2)+(0, 0.85)$) {$C_2$};
                
                \draw[thick] ($(c7)+(0,0.65)$) -- ($(c7)+(0,1.15)$) -- ($(c12)+(0,1.15)$) -- ($(c12)+(0,0.65)$);
                \node at ($(c9)+(0.5,1.45)$) {$\tilde{C}$};
            \end{scope}
            
            \begin{scope}
                \draw[rounded corners] (0, -0.5) rectangle (8, 0.5);
                
                \node at (-0.5, 0) {$H$};
                \foreach \i in {1,...,8} {
                    \pgfmathsetmacro{\x}{\i-0.5}
                    \Vertex[x=\x, y=0]{h\i}
                }
                \Edge(h1)(c7)
                \Edge(h2)(c8)
                \Edge(h3)(c9)
                \Edge(h4)(c10)
                \Edge(h5)(c11)
                \Edge(h6)(c12)
                \Edge(h7)(c13)
                \Edge(h8)(c14)
                \Edges(c4,h1,c6)
                \Edges(h2,c5)
            \end{scope}
    
            \begin{scope}[on background layer]

                \fill[goodteal, rounded corners] ($(h1)-(0.4, 0.4)$) rectangle ($(h4)+(0.4, 0.4)$);
                \node at ($(h2)+(0.5, 0.85)$) {$F$};
                
                \fill[goodgreen, rounded corners] ($(h1)-(0.3, 0.3)$) rectangle ($(h2)+(0.3, 0.3)$);
                \node at ($(h1)+(0.5, -0.85)$) {$X_0$};

                \draw[thick] ($(h1)-(0,0.65)$) -- ($(h1)-(0,1.15)$) -- ($(h6)-(0,1.15)$) -- ($(h6)-(0,0.65)$);
                \node at ($(h4)-(0.5,1.45)$) {$\tilde{H}$};

                \fill[goodyellow, rounded corners] ($(h5)-(0.3, 0.3)$) rectangle ($(c12)+(0.3, 0.3)$);
                \node at ($(h5)+(0.5,1.5)$) {$G'$};
                
                \fill[goodolive, rounded corners] ($(h7)-(0.3, 0.3)$) rectangle ($(h8)+(0.3, 0.3)$);
            \end{scope}

            
        
        \end{tikzpicture}
        \caption{Example of the structures involved in the algorithm $\propLarge$, used in the proof of \cref{lemma:SmallCrownBigCrown}. The five light-blue-shaded vertices correspond to the inputs $C_1, C_2$ of $\propLarge$; the two olive-shaded vertices are forced to be in the solution by our choice of $C_1$. $\bar{X}_1$, i.e., the set of vertices of $C \setminus V(M)$ that should be forbidden in a solution to $G$, is represented by the three red-shaded vertices; the dark-green shaded vertices are forced to be in every solution that avoids $\bar{X}_1$, and, together with $\tilde{G} = G[\tilde{H} \cup \tilde{C}]$, composes the input to the call to $\propDir$ used in $\propLarge$; the four teal-shaded and four purple-shaded vertices are the outputs $F,\bar{F}$ of $\propDir$, and, together with $\bar{X}_1$, make up the output of $\propLarge$. Finally, the four yellow-shaded vertices represent $G'$, the instance of \pname{Enum Small Crown} to which $\A_s$ will be applied.}
        \label{fig:propagateC1C2}
    \end{figure}
    
    Our next step is guessing which edges of the $H$-saturating matching $M$ will have both endpoints in the solution $S$; in the next propagation procedure, this is represented by set $C_1$, i.e. which vertices of $C \cap V(M)$ must be accompanied by their image in $S$.
    We also guess which vertices of $C$ that \emph{are not} matched to $H$ will be included in a solution; this is represented by $C_2$.
    Finally, the choice of $C_1$ and $C_2$ implies that the vertices $\bar{X} \subseteq C \setminus (V(M) \cup C_2)$ are incident to yet uncovered edges.
    Formally, for any $d \in  [\max(0,x-(|C|-|V(M))|),x]$, any set $C_1 \subseteq V(M) \cap C$ of size $d$, and any set $C_2 \subseteq C \setminus V(M)$ of size $x-d$,
    we define (see \cref{fig:propagateC1C2}) $\propLarge(C_1,C_2)$ that computes $(F,\bar{F},\bar{X_1})$ where $\bar{X_1}=(C \setminus V(M)) \setminus C_2$,
    and $(F,\bar{F})=\propDir(\tilde{G},X_0)$ with $X_0 = N(\bar{X_1})$, and $\tilde{G} =G[\tilde{H} \cup \tilde{C}]$, where $\tilde{H} = H \setminus M_V(C_1)$, and $\tilde{C}=M_V(\tilde{H})$.
    
    \begin{claim}\label{claim:sign}
        For any $S \in \Sol(G,x)$ of signature $(C_1,C_2)$,
        \begin{enumerate}
            \item \label{enum:prop1} $S$ contains $F$ and avoids $\bar{F} \cup \bar{X_1}$;
            \item \label{enum:prop2} By defining $H'= \tilde{H} \setminus  F$ and $C'= \tilde{C} \setminus \bar{F}$,
            $G' = G[H' \cup C']$ is a small crowned graph and $(S \cap (C' \cup H')) \in \Sol(G')$.
        \end{enumerate}
    \end{claim}
    
    \begin{proof}
        Let $S \in \Sol(G,x)$ have signature $(C_1,C_2)$.
        By definition of $C_2(S)$, we get that $S \cap \bar{X_1} = \emptyset$, implying also that $X_0 \subseteq S$.
        Moreover, by definition of $C_1(S)$, no edge $e \in M$ with $V(e) \subseteq \tilde{H} \cup \tilde{C}$ has both its endpoints in $S$.
        Thus, if we define $\tilde{S}= S \cap (\tilde{H} \cup \tilde{C})$, we get that $\tilde{S} \in \Sol(\tilde{G})$ and that $X_0 \subseteq \tilde{S}$.
        By \cref{lemma:propagateAux}, we obtain the two claimed properties.
    \end{proof}

    We are now ready to define $\A$. To this end, let $\A_s$ be the polynomial-delay algorithm for \pname{Enum Small Crown}.
    For any $d \in [\max(0,x-(|C|-|V(M)|)),x]$, any set $C_1 \subseteq V(M) \cap C$ of size $d$, and any set $C_2 \subseteq C \setminus V(M)$ of size $x-d$,
    $\A$ runs $\propLarge(C_1,C_2)$ that computes $(F,\bar{F},\bar{X_1})$, and defines $G'$, $H'$, and $C'$ as in \cref{enum:prop2} of \cref{claim:sign}. Note that $G'$ may be the empty graph.
    Then, $\A$ outputs all solutions of the form $C_1 \cup M(C_1) \cup C_2 \cup F \cup S'$, where $S' \in \Sol(G')$ are enumerated by $\A_s$.
    Observe that we can indeed call $\A_s$ on $G'$ as according to \cref{enum:prop2}, $G'$ is a small crowned graph.

    Let us first prove that there is no repetition. Observe that for each $d$, $C_1$, $C_2$ considered by $\A$, all enumerated solutions of the form
    $C_1 \cup M(C_1) \cup C_2 \cup F \cup S'$ (where $S' \in \Sol(G')$) have a signature $(C_1,C_2)$. 
    Now, it is sufficient to observe that two solutions of $(G,x)$ with different signatures are necessarily different, which implies that there is no repetition.
    
    Let us now prove that all enumerated elements are in $\Sol(G,x)$.
    Let $S=C_1 \cup M(C_1) \cup C_2 \cup F \cup S'$, where $S' \in \Sol(G')$ exists (as $S' = S \cap V(G')$ and $G'$ is an induced subgraph of $G$) and is enumerated by $\A$.
    By \cref{aux:prop3} of \cref{lemma:propagateAux}, $(S' \cup F) \in \Sol(\tilde{G})$ and contains $X_0$.
    Now, observe that $H \cup C$ is partitioned into $V_1=C_1 \cup M(C_1) \cup C_2 \cup \bar{X_1}$, and $V_2=\tilde{H} \cup \tilde{C}$.
    Notice that $S \cap V_i$ is a vertex cover of $G[V_i]$ for $i \in \{1,2\}$. Moreover, as $V_1 \setminus S = \bar{X_1}$, the only edges between $V_1$ and $V_2$
    not covered by $S \cap V_1$ are edges between $\bar{X_1}$ and $N(\bar{X_1})=X_0$, but as $S \cap V_2 \supseteq X_0$, $S$ is indeed a vertex cover of $G$.
    Finally, $|S|=2|C_1|+|C_2|+|F|+|H'|=2|C_1|+|C_2|+|\tilde{H}|=|H|+|C_1|+|C_2|=|H|+x$, and so we have $S \in \Sol(G,x)$.
    
    Towards proving that all elements of $\Sol(G,x)$ are enumerated, let $S \in \Sol(G,x)$, and $(C_1(S),C_2(S))$ be its signature.
    By definition of $\A$, there exists $C_1$ and $C_2$ considered by $\A$ such that $C_i=C_i(S)$ for $i \in \{1,2\}$. 
    Let $(F,\bar{F},\bar{X_1})=\propLarge(C_1,C_2)$. 
    By \cref{enum:prop1}, $S \supseteq F$, and $S \cap (\bar{F} \cup \bar{X_1}) = \emptyset$, which implies that
    $S=C_1 \cup M(C_1) \cup C_2 \cup F \cup S'$. As $S \cap (C' \cup H') = S'$, \cref{enum:prop2} implies that $S' \in \Sol(G')$, and so $S'$ is enumerated by $\A_s$.
    
    It remains to prove that there is only polynomial-time between two consecutive outputs of $\A$. Notice first that $\propLarge$ runs in polynomial time.
    Then, for each fixed $C_1,C_2$ considered by $\A$,  as $\A_s$ runs in polynomial-delay, there is at most polynomial-time between two consecutive solutions having the same signature $(C_1,C_2)$.
    Moreover, and as for any choice of $C_1,C_2$ considered by $\A$, we know that $\Sol(G') \neq \emptyset$ (by \cref{obs:enumCrown}), there is also a polynomial-delay between two consecutive solutions with different signatures.
\end{proof}

The proof of \cref{lemma:SmallCrown} requires a more involved variant of the propagation algorithm, where the goal is to avoid a prescribed vertex~$v$.

\begin{figure}[!htb]
    \centering
    \begin{subfigure}[t]{0.5\textwidth}
        \begin{tikzpicture}[scale=0.8]
            \GraphInit[unit=3,vstyle=Normal]
            \SetVertexNormal[Shape=circle, FillColor=black, MinSize=2pt]
            \tikzset{VertexStyle/.append style = {inner sep = \inners, outer sep = \outers}}
            \SetVertexNoLabel
            \SetUpEdge[style={opacity=0.3}]
            
            \newcommand{\myncount}{7}
            \begin{scope}
                \draw[rounded corners] (0, -0.5) rectangle (\myncount, 0.5);
                
                \node at (-0.5, 0) {$H$};
                \foreach \i in {1,...,\myncount} {
                    \pgfmathsetmacro{\x}{\i-0.5}
                    \Vertex[x=\x, y=0]{h\i}
                }
            \end{scope}
            \begin{scope}[yshift=3cm]
                \draw[dashed, rounded corners] (0, -0.5) rectangle (\myncount, 0.5);
                
                \node at (-0.5, 0) {$C$};
                \foreach \i in {1,...,\myncount} {
                    \pgfmathsetmacro{\x}{\i-0.5}
                    \Vertex[x=\x, y=0]{c\i}
                    \Edge(h\i)(c\i)
                }
            \end{scope}
            
            \Edge[style={-Latex, bend right}](h1)(h3)
            \Edge[style={-Latex}](h1)(c2)
            \Edge[style={-Latex}](h1)(c1)
            \Edge(h2)(h3)
            \Edges(h4,c3,h5)
            \Edges(c5,h3,c6)
            
            \Edge[style={bend right,opacity=0.3}](h5)(h7)
            \Edges(h6,h7,c7,h6)

            \begin{scope}[on background layer]
              \fill[goodpurple, rounded corners] ($(h1)-(0.3, 0.3)$) rectangle ($(h1)+(0.3, 0.3)$);
              
              \fill[goodcyan, rounded corners] ($(c1)-(0.3, 0.3)$) rectangle ($(c2)+(0.3, 0.3)$);
              \fill[goodcyan, rounded corners] ($(h3)-(0.3, 0.3)$) rectangle ($(h3)+(0.3, 0.3)$);
            \end{scope}
        \end{tikzpicture}
        \subcaption{Computing $F_1$ (cyan) from $\bar{F}_0$ (purple).}
    \end{subfigure}%
    ~
    \begin{subfigure}[t]{0.5\textwidth}
        \begin{tikzpicture}[scale=0.8]
            \GraphInit[unit=3,vstyle=Normal]
            \SetVertexNormal[Shape=circle, FillColor=black, MinSize=2pt]
            \tikzset{VertexStyle/.append style = {inner sep = \inners, outer sep = \outers}}
            \SetVertexNoLabel
            \SetUpEdge[style={opacity=0.3}]
            \newcommand{\myncount}{7}
            \begin{scope}
                \draw[rounded corners] (0, -0.5) rectangle (\myncount, 0.5);
                
                \node at (-0.5, 0) {$H$};
                \foreach \i in {1,...,\myncount} {
                    \pgfmathsetmacro{\x}{\i-0.5}
                    \Vertex[x=\x, y=0]{h\i}
                }
            \end{scope}
            \begin{scope}[yshift=3cm]
                \draw[dashed, rounded corners] (0, -0.5) rectangle (\myncount, 0.5);
                
                \node at (-0.5, 0) {$C$};
                \foreach \i in {1,...,\myncount} {
                    \pgfmathsetmacro{\x}{\i-0.5}
                    \Vertex[x=\x, y=0]{c\i}
                    \Edge(h\i)(c\i)
                }
            \end{scope}
            
            \Edge[style={bend right}](h1)(h3)
            \Edge(h1)(c2)
            \Edge(h1)(c1)
            \Edge(h2)(h3)
            \Edges(h4,c3,h5)
            \Edges(c5,h3,c6)
            
            \Edge[style={bend right,opacity=0.3}](h5)(h7)
            \Edges(h6,h7,c7,h6)
            
            \Edge[style={-Latex}](c1)(h1)
            \Edge[style={-Latex}](c2)(h2)
            \Edge[style={-Latex}](h3)(c3)

            \begin{scope}[on background layer]
              \fill[goodred, rounded corners] ($(h1)-(0.3, 0.3)$) rectangle ($(h2)+(0.3, 0.3)$);
              \fill[goodred, rounded corners] ($(c3)-(0.3, 0.3)$) rectangle ($(c3)+(0.3, 0.3)$);
              
              \fill[goodteal, rounded corners] ($(c1)-(0.3, 0.3)$) rectangle ($(c2)+(0.3, 0.3)$);
              \fill[goodteal, rounded corners] ($(h3)-(0.3, 0.3)$) rectangle ($(h3)+(0.3, 0.3)$);
            \end{scope}
        \end{tikzpicture}
        \subcaption{Computing $\bar{F}_1$ (red) from $F_1$ (teal).}
    \end{subfigure}%

    \vspace{1cm}
    \begin{subfigure}[t]{0.5\textwidth}
        \begin{tikzpicture}[scale=0.8]
            \GraphInit[unit=3,vstyle=Normal]
            \SetVertexNormal[Shape=circle, FillColor=black, MinSize=2pt]
            \tikzset{VertexStyle/.append style = {inner sep = \inners, outer sep = \outers}}
            \SetVertexNoLabel
            \SetUpEdge[style={opacity=0.3}]
            
            \newcommand{\myncount}{7}
            \begin{scope}
                \draw[rounded corners] (0, -0.5) rectangle (\myncount, 0.5);
                
                \node at (-0.5, 0) {$H$};
                \foreach \i in {1,...,\myncount} {
                    \pgfmathsetmacro{\x}{\i-0.5}
                    \Vertex[x=\x, y=0]{h\i}
                }
            \end{scope}
            \begin{scope}[yshift=3cm]
                \draw[dashed, rounded corners] (0, -0.5) rectangle (\myncount, 0.5);
                
                \node at (-0.5, 0) {$C$};
                \foreach \i in {1,...,\myncount} {
                    \pgfmathsetmacro{\x}{\i-0.5}
                    \Vertex[x=\x, y=0]{c\i}
                    \Edge(h\i)(c\i)
                }
            \end{scope}
            
            \Edge(h1)(c2)
            \Edge(h1)(c1)
            \Edge(h2)(h3)
            \Edges(h4,c3,h5)
            \Edges(c5,h3,c6)
            
            \Edge[style={bend right,opacity=0.3}](h5)(h7)
            \Edges(h6,h7,c7,h6)

            \Edge[style={-Latex, bend right}](h1)(h3)
            \Edge[style={-Latex}](h1)(c2)
            \Edge[style={-Latex}](h2)(c2)
            \Edge[style={-Latex}](h1)(c1)
            \Edge[style={-Latex}](c1)(h1)
            \Edge[style={-Latex}](c3)(h3)
            \Edge[style={-Latex}](c3)(h4)
            \Edge[style={-Latex}](c3)(h5)

            \begin{scope}[on background layer]
              \fill[goodpurple, rounded corners] ($(h1)-(0.3, 0.3)$) rectangle ($(h2)+(0.3, 0.3)$);
              \fill[goodpurple, rounded corners] ($(c3)-(0.3, 0.3)$) rectangle ($(c3)+(0.3, 0.3)$);
              
              \fill[goodcyan, rounded corners] ($(c1)-(0.3, 0.3)$) rectangle ($(c2)+(0.3, 0.3)$);
              \fill[goodcyan, rounded corners] ($(h3)-(0.3, 0.3)$) rectangle ($(h5)+(0.3, 0.3)$);
            \end{scope}
        \end{tikzpicture}
        \subcaption{Computing $F_2$ (cyan) from $\bar{F}_1$ (purple).}
    \end{subfigure}%
    ~
    \begin{subfigure}[t]{0.5\textwidth}
        \begin{tikzpicture}[scale=0.8]
            \GraphInit[unit=3,vstyle=Normal]
            \SetVertexNormal[Shape=circle, FillColor=black, MinSize=2pt]
            \tikzset{VertexStyle/.append style = {inner sep = \inners, outer sep = \outers}}
            \SetVertexNoLabel
            \SetUpEdge[style={opacity=0.3}]
            
            \newcommand{\myncount}{7}
            \begin{scope}
                \draw[rounded corners] (0, -0.5) rectangle (\myncount, 0.5);
                
                \node at (-0.5, 0) {$H$};
                \foreach \i in {1,...,\myncount} {
                    \pgfmathsetmacro{\x}{\i-0.5}
                    \Vertex[x=\x, y=0]{h\i}
                }
            \end{scope}
            \begin{scope}[yshift=3cm]
                \draw[dashed, rounded corners] (0, -0.5) rectangle (\myncount, 0.5);
                
                \node at (-0.5, 0) {$C$};
                \foreach \i in {1,...,\myncount} {
                    \pgfmathsetmacro{\x}{\i-0.5}
                    \Vertex[x=\x, y=0]{c\i}
                    \Edge(h\i)(c\i)
                }
            \end{scope}
            
            \Edge[style={bend right}](h1)(h3)
            \Edge(h1)(c2)
            \Edge(h1)(c1)
            \Edge(h2)(h3)
            \Edges(h4,c3,h5)
            \Edges(c5,h3,c6)
            
            \Edge[style={bend right,opacity=0.3}](h5)(h7)
            \Edges(h6,h7,c7,h6)

            \Edge[style={bend right}](h1)(h3)
            \Edge[style={-Latex}](c1)(h1)
            \Edge[style={-Latex}](c2)(h2)
            \Edge[style={-Latex}](h3)(c3)
            \Edge[style={-Latex}](h4)(c4)
            \Edge[style={-Latex}](h5)(c5)

            \begin{scope}[on background layer]
              \fill[goodred, rounded corners] ($(h1)-(0.3, 0.3)$) rectangle ($(h2)+(0.3, 0.3)$);
              \fill[goodred, rounded corners] ($(c3)-(0.3, 0.3)$) rectangle ($(c5)+(0.3, 0.3)$);
              
              \fill[goodteal, rounded corners] ($(c1)-(0.3, 0.3)$) rectangle ($(c2)+(0.3, 0.3)$);
              \fill[goodteal, rounded corners] ($(h3)-(0.3, 0.3)$) rectangle ($(h5)+(0.3, 0.3)$);
            \end{scope}
        \end{tikzpicture}
        \subcaption{Computing $\bar{F}_2$ (red) from $F_2$ (teal).}
    \end{subfigure}

    \vspace{1cm}
    \begin{subfigure}[t]{0.5\textwidth}
        \begin{tikzpicture}[scale=0.8]
            \GraphInit[unit=3,vstyle=Normal]
            \SetVertexNormal[Shape=circle, FillColor=black, MinSize=2pt]
            \tikzset{VertexStyle/.append style = {inner sep = \inners, outer sep = \outers}}
            \SetVertexNoLabel
            \SetUpEdge[style={opacity=0.3}]
            \newcommand{\myncount}{7}
            \begin{scope}
                \draw[rounded corners] (0, -0.5) rectangle (\myncount, 0.5);
                
                \node at (-0.5, 0) {$H$};
                \foreach \i in {1,...,\myncount} {
                    \pgfmathsetmacro{\x}{\i-0.5}
                    \Vertex[x=\x, y=0]{h\i}
                }
            \end{scope}
            \begin{scope}[yshift=3cm]
                \draw[dashed, rounded corners] (0, -0.5) rectangle (\myncount, 0.5);
                
                \node at (-0.5, 0) {$C$};
                \foreach \i in {1,...,\myncount} {
                    \pgfmathsetmacro{\x}{\i-0.5}
                    \Vertex[x=\x, y=0]{c\i}
                    \Edge(h\i)(c\i)
                }
            \end{scope}
            
            \Edge[style={-Latex, bend right}](h1)(h3)
            \Edge[style={-Latex}](h1)(c2)
            \Edge[style={-Latex}](h1)(c1)
            \Edge[style={-Latex}](h1)(c3)
            \Edge(h2)(h3)
            \Edges(h4,c3,h5)
            \Edges(c5,h3,c6)
            
            \Edge[style={bend right,opacity=0.3}](h5)(h7)
            \Edges(h6,h7)
            \Edges(c7,h6)
            \SetUpEdge[color=goodolive,style={line width=2pt}]
            \Edges(h3,c3)

            \begin{scope}[on background layer]
              \fill[goodpurple, rounded corners] ($(h1)-(0.3, 0.3)$) rectangle ($(h1)+(0.3, 0.3)$);
              
              \fill[goodcyan, rounded corners] ($(c1)-(0.3, 0.3)$) rectangle ($(c3)+(0.3, 0.3)$);
              \fill[goodcyan, rounded corners] ($(h3)-(0.3, 0.3)$) rectangle ($(h3)+(0.3, 0.3)$);
            \end{scope}
        \end{tikzpicture}
        \subcaption{Failing to compute $F_1$ (cyan) from $\bar{F}_0$ (purple).}
    \end{subfigure}%
    ~
    \begin{subfigure}[t]{0.5\textwidth}
        \begin{tikzpicture}[scale=0.8]
            \GraphInit[unit=3,vstyle=Normal]
            \SetVertexNormal[Shape=circle, FillColor=black, MinSize=2pt]
            \tikzset{VertexStyle/.append style = {inner sep = \inners, outer sep = \outers}}
            \SetVertexNoLabel
            \newcommand{\myncount}{7}
            \SetUpEdge[style={opacity=0.3}]
            \begin{scope}
                \draw[rounded corners] (0, -0.5) rectangle (\myncount, 0.5);
                
                \node at (-0.5, 0) {$H$};
                \foreach \i in {1,...,\myncount} {
                    \pgfmathsetmacro{\x}{\i-0.5}
                    \Vertex[x=\x, y=0]{h\i}
                }
            \end{scope}
            \begin{scope}[yshift=3cm]
                \draw[dashed, rounded corners] (0, -0.5) rectangle (\myncount, 0.5);
                
                \node at (-0.5, 0) {$C$};
                \foreach \i in {1,...,\myncount} {
                    \pgfmathsetmacro{\x}{\i-0.5}
                    \Vertex[x=\x, y=0]{c\i}
                    \Edge(h\i)(c\i)
                }
            \end{scope}
            
            \Edge(h1)(c2)
            \Edge(h1)(c1)
            \Edges(h4,c3,h5)
            \Edges(c5,h3,c6)
            
            \Edge[style={bend right,opacity=0.3}](h5)(h7)
            \Edges(h6,h7,c7,h6)
            
            \Edge[style={-Latex}](c1)(h1)
            \Edge[style={-Latex}](c2)(h2)
            \Edge[style={-Latex}](c3)(h3)
            \Edge(h1)(c3)

            \begin{scope}[on background layer]
              \fill[goodred, rounded corners] ($(h1)-(0.3, 0.3)$) rectangle ($(h3)+(0.3, 0.3)$);
              
              \fill[goodteal, rounded corners] ($(c1)-(0.3, 0.3)$) rectangle ($(c3)+(0.3, 0.3)$);
            \end{scope}

            \SetUpEdge[style={line width=2pt, color=goodyellow}]
            \Edge(h2)(h3)
            \Edge[style={bend right,line width=2pt, color=goodyellow}](h1)(h3)
        \end{tikzpicture}
        \subcaption{Failing to compute $\bar{F}_1$ (red) from $F_1$ (teal).}
    \end{subfigure}%
    \caption{Examples of executions of algorithm $\propAll$. Vertical edges correspond to $M$.
    We reduce the width of the edges of $G$ not involved in the corresponding operation for clarity.
    An arc $(x,y)$ indicates that vertex $y$ was included in the set currently being built by virtue of the membership of $x$ in the root set.
    The edges that cause the failures in the computation of $F_i$ or $\bar{F}_i$ are thickened and colored differently. \label{fig:propagateAuxBar}}
\end{figure}
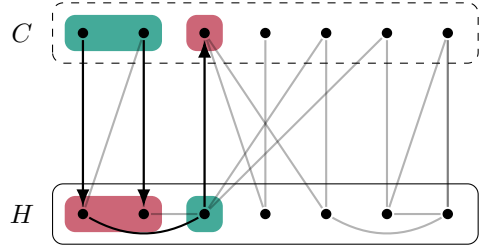
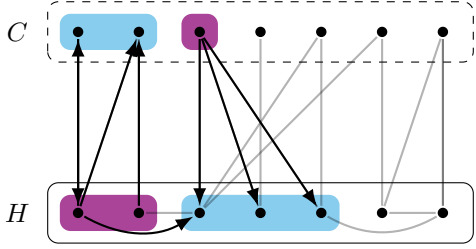
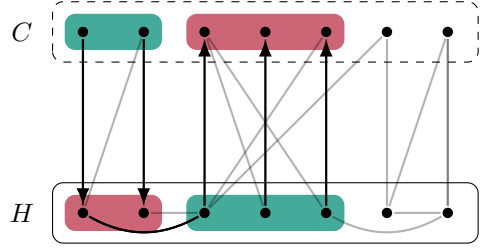
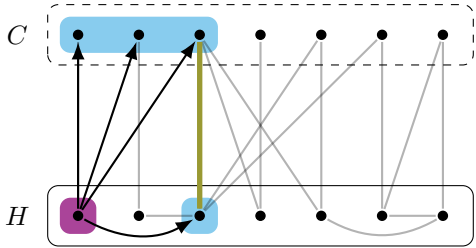
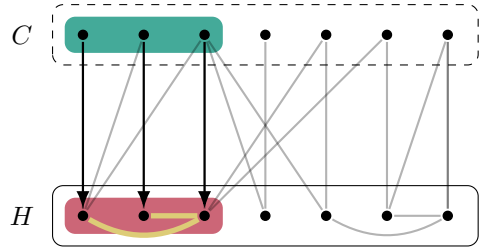

\begin{definition}[$\propAll(G,v)$]\label{prop-avoid-v}
    Given a small crowned graph $G = (H \cup C, E)$ (i.e., $|H|=|C|$), and $v \in H$, the procedure $\propAll(G,v)$ either \emph{fails},
    or outputs two sets $(F,\bar{F})$ that, intuitively, are forced to be included or avoided in a solution that avoids $v$ and, upon removal, will yield a small crowned graph. We recursively define them as follows.
    Let $\bar{F_0}=\{v\}$. Observe that $N(\bar{F_0})$ intersects $C$ (because of the perfect matching $M$ between $H$ and $C$, and may intersect $H$).
    Now, for any $i \geq 1$ (see \cref{fig:propagateAuxBar}):
    \begin{itemize}
        \item to compute $F_{i}$: if there exists $e \in M$ such that $V(e) \subseteq N(\bar{F}_{i-1})$, we fail, and otherwise we define $F_{i}=N(\bar{F}_{i-1})$.
        \item to compute $\bar{F}_i$: if $M(F_i)$ is not an independent set, we fail, and otherwise we define $\bar{F_i}=M(F_i)$; note that $\bar{F}_{i-1} \subseteq \bar{F}_i$.
    \end{itemize}
    We continue this process until it fails or until $F_i = F_{i-1}$ and $\bar{F}_i = \bar{F}_{i-1}$.
    If there is no failure, $\propAll(G,v)$ returns $(F, \bar{F}) = (F_i, \bar{F_i}$).
\end{definition}

\begin{lemma}\label{lemma:propagateAuxBar}
    Let $G = (H \cup C, E)$ be a small crowned graph, and $v \in H$.
    \begin{enumerate}
        \item \label{auxBar:prop1} If $\propAll(G,v)$ fails, then there is no $S \in \Sol(H,C)$ such that $v \notin S$.
        \item \label{auxBar:prop2} Otherwise, if $\propAll(G,v)$ returns $(F,\bar{F})$ (where $\bar{F}=M_V(F)$), then for any $S \in \Sol(G)$ such that $v \notin S$,
        $F \subseteq S$ and $S \cap \bar{F} = \emptyset$; moreover, if we define $H'= H \setminus (F \cup \bar{F})$ and $C'= C \setminus (F \cup \bar{F})$,
        then $G' = G[H' \cup C']$ is a crowned graph with $|C'|=|H'|$, and $S \cap (C' \cup H') \in \Sol(G')$.
        \item \label{auxBar:prop3} For any $S' \in \Sol(G')$, $F \cup S'$ is a solution of $G$ that avoids $v$.
    \end{enumerate}
\end{lemma}

\begin{proof}
  Suppose that there is some $S \in \Sol(G)$ where $v \notin S$.
  Let us prove by induction that for any $i \geq 1$, if there is a failure when computing $F_i$ or $\bar{F_i}$, then $S$ cannot exist,
  and otherwise $F_i \subseteq S$ and $S \cap \bar{F_i} = \emptyset$.
  
  Assume there was no failure, and consider the step where we try to compute $F_i$.
  By induction, $S \cap \bar{F}_{i-1} = \emptyset$, so we have that $N(\bar{F}_{i-1}) \subseteq S$ and, by construction, $F_i = N(\bar{F}_{i-1})$.
  By \cref{obs:enumCrown}, if there is an edge $e \in M$ where $V(e) \subseteq F_i$,
  then no such solution exists. Let us now consider the case where we try to compute $\bar{F_i}$.
  Again by induction, we have $F_{i-1} \subseteq S$, and by \cref{obs:enumCrown}
  we know that no edge of $M$ can have both endpoints in $S$, implying that we must have $S \cap M_V(F_i) = \emptyset$.
  Thus, if $G[M_V(F_i)]$ is not an independent set, then no such solution exists.

  The correctness of \cref{auxBar:prop1} now directly follows from the induction.
  The proof of \cref{auxBar:prop2} is also immediate: if $\propAll(G,v)$ returns $(F,\bar{F})$, then $\bar{F}=M_V(F)$ follows from the definition, and for
  $S \in \Sol(G)$ such that $v \notin S$, $F \subseteq S$ and $S \cap \bar{F} = \emptyset$ follows from the induction, and the fact that $(S \cap (C' \cup H')) \in \Sol(G')$ is true by \cref{obs:enumCrown}.

  To prove \cref{auxBar:prop3}, let $S' \in \Sol(G')$, and let $S=F \cup S'$.
  Notice that $H \cup C$ is partitioned into $F \cup \bar{F}$ and $H' \cup C'$, that $F$ is a vertex cover of $G[F \cup \bar{F}]$ (as $\bar{F} = M_V(F)$),
  and that $S'$ is a vertex cover of $G[H' \cup C']$.
  Now, let $i_0$ be the index where $\propAll$ stops, meaning that $F_{i_0} = F_{i_0-1}$.
  This implies that $N(\bar{F}_{i_0}) \subseteq F_{i_0-1}$,
  and thus that there is no edge between $\bar{F}_{i_0}=\bar{F}$ and $H' \cup C'$, implying that $S$ is a vertex cover of $G[H \cup C]$.
  As $|S|=|H|$, we get that $S$ is a solution of $G$ that avoids $v$ by definition of $F$.
\end{proof}

With the propagation algorithm in hand, \cref{lemma:SmallCrown} can be proved by constructing a branching algorithm where we either add a vertex $v$ of $H$ or do not add it to the solution.
The key properties we use are that: (i) there is always one solution that uses $v$ (e.g. taking the entirety of $H$), and (ii) we can decide in polynomial time if it is possible to avoid $v$ in a solution. Consequently, we recursively call our branching algorithm only if we know that there is some solution to be found in this branch.

\begin{restatable}{lemma}{lemmaSmallCrown}{}
    \label{lemma:SmallCrown}
    There is a polynomial-delay algorithm for \pname{Enum Small Crown}.
\end{restatable}

\begin{proof}
    Let $G = (H \cup C, E)$ be a small crowned graph and let us define an algorithm $\A_s$ that enumerates all vertex covers of $G$ of size $|H|$.
    If $V(G) = \emptyset$, then $\A_s$ returns $\emptyset$, so now let $v \in H$ be an arbitrary vertex.
    Let $(F_v,\bar{F_v})=\propDir(G,\{v\})$, $H'_v=H \setminus F_v$, $C'_v = C \setminus \bar{F_v}$, and $G'_v = G[H'_v \cup C'_v]$.
    Algorithm $\A_s$ outputs first all solutions of the form $F_v \cup S'$, where $S'$ are enumerated by a recursive call $\A_s(G'_v)$.
    Then, $\A_s$ calls $\propAll(G,v)$. If it fails then $\A_s$ stops, otherwise let $(F_{\bar{v}},\bar{F}_{\bar{v}})$ be the output of $\propAll(G,\bar{v})$, $H'_{\bar{v}}=H \setminus F_{\bar{v}}$, $C'_{\bar{v}} = C \setminus \bar{F}_{\bar{v}}$, and $G'_{\bar{v}} = G[H'_{\bar{v}} \cup C'_{\bar{v}}]$.
    Algorithm $\A_s$ outputs then all solutions of the form $F_{\bar{v}} \cup S'$, where $S'$ are enumerated by a recursive call
    $\A_S(G'_{\bar{v}})$.
    This concludes the definition of $\A_s$.
    
    First, observe that the recursive calls are made on small crowned graphs, as they were obtained by removing sets $(F,\bar{F})$ from $V(G)$
    where $\bar{F}=M_V(F)$.
    The fact that there is no repetition is immediate by induction, as in particular all solutions of $F_v \cup S'$, where $S' \in \Sol(G'_v)$, contain $v$,
    and all solutions of the form $F_{\bar{v}} \cup S''$, where $S'' \in \Sol(G'_{\bar{v}})$, do not contain $v$.

    Let us prove by induction that all enumerated elements are in $\Sol(G)$.
    We first consider enumerated sets of the form  $F_v \cup S'$, where $S'$ are enumerated by $\A_s(G'_v)$. By induction, we get that $S' \in \Sol(G'_v)$,
    and by \cref{aux:prop3} of \cref{lemma:propagateAux}, we get that $F_v \cup S' \in \Sol(G)$.
    Now, for enumerated sets of the form $F_{\bar{v}} \cup S''$, where $S''$ are enumerated by $\A_s(G'_{\bar{v}})$, the induction hypothesis implies that
    $S' \in \Sol(G'_{\bar{v}})$, and again by \cref{auxBar:prop3} of \cref{lemma:propagateAuxBar}, we have that $F_{\bar{v}} \cup S'' \in \Sol(G)$.

    Let us now prove by induction that all elements of $\Sol(G)$ are enumerated.
    Let $S \in \Sol(G)$.
    If $v \in S$, then by \cref{lemma:propagateAux}, $S=F_v \cup S'$, where $S' \in \Sol(G'_v)$, and by induction $S'$ will be enumerated by $\A_s(G'_v)$.
    If $v \notin S$, then by \cref{lemma:propagateAuxBar}, as $\propAll(G,v)$ cannot fail (as $S$ exists), $S=F_{\bar{v}} \cup S'$, where $S' \in \Sol(H'_{\bar{v}},C'_{\bar{v}})$,  and by induction $S'$ will be enumerated by $\A_s(H'_{\bar{v}},C'_{\bar{v}})$.
    
    Finally, there is at most polynomial time between any two solutions, as both $\propDir$ and $\propAll$ run in polynomial time, and because for any recursive call
    made on an input $G' \in \{G'_v, G'_{\bar{v}}\}$, we know according to \cref{obs:enumCrown} that $\Sol(G') \neq \emptyset$.
    Moreover, the branching tree has height bounded by $|V(G)|$ and every node either: is a leaf, and so falls in the base case where $V(G) = \emptyset$; has exactly one child, corresponding to $\A_s(G'_v)$, for which a solution always exists; or has two children where $\A_s(G'_v)$ has the same guarantee as in the previous case, and $\A_s(G'_{\bar{v}})$ is called if and only if we know that $\Sol(G'_{\bar{v}}) \neq \emptyset$ by \cref{lemma:propagateAuxBar}.
    Consequently, every leaf of the branching tree corresponds to a solution of $G$ and the time taken between visiting two of them is bounded by a polynomial in $|V(G)| + |E(G)|$, guaranteeing the desired polynomial delay.
\end{proof}

The following theorem is now a direct consequence of \cref{lemma:crownImpliesKernel,lemma:SmallCrownBigCrown,lemma:SmallCrown}.

\thmvcvcstronglinear*

\section{Feedback Vertex Set}
\label{sec:fvs}

Throughout this section, we take $(G, k)$ to be our input instance to \pname{Enum Feedback Vertex Set}, let $X \subseteq V(G)$ be a 2-approximation using the algorithm given in~\cite{bafna_fvs_apx}, $\fvs(G)$ denote the size of a minimum feedback vertex set, and $\Delta(G)$ its maximum degree.
Thomassé's~\cite{thomasse_fvs} strategy can be split into four phases: (i) apply reduction rules to bound the maximum and minimum degrees of the instance; (ii) prove that $|V(G)| \in \bigO{\Delta(G) \cdot \fvs(G)}$ in graphs of minimum degree three; (iii) identify if there is a vertex $v$ in $X$ that belong to too many vertex-disjoint cycles in $(G \setminus X) \cup \{v\}$; and (iv) once no such $v$ exists, pick a high-degree vertex $u$, compute a $2$-expansion (i.e. a crown decomposition where each head vertex is matched to two unique crown vertices) in an auxiliary graph, and remove all edges from $u$ to the vertices identified by the expansion.
As exhaustively applying (iv) implies a bound on the maximum degree, (i) and (ii) will yield the desired kernel.
We follow a similar kernelization strategy, but use a substantially different set of rules beyond the basic ones; our presentation is closer to that of Fomin et al.~\cite[Chapter 5]{book_kernels}.
We present a generalization of point (i) in \cref{lem:fvsfvs_bounding_lemma}: $|V(G)| \in \bigO{\Delta(G) \cdot \fvs(G)}$ holds even in some of graphs of minimum degree two.

We begin by bounding the minimum degree of $G$; like other works dealing with multigraphs, we define $\deg_G(v)$ as the number of edges incident to a vertex $v$ in $G$, and omit the subscript when $G$ is clear from the context.
We say that $uv \in E(G)$ is a \textit{double-edge} if the multiplicity of $uv$ in $E(G)$, denoted by $\mul(uv)$, is two.
Note that a double-edge is equivalent to a cycle of length two in $G$ or, equivalently, that at least one of its endpoints must be present in \emph{every} solution of the instance.
Our first rule is only used to simplify the graph.

\begin{restatable}{fvsfvsrule}{rrulefvsfvseasycases}{}
    \label{rrule:fvsfvs_easy_cases}
    Perform the following modifications:
    \begin{enumerate}[i.]
        \item\label{item:fvsfvs_easy_cases1} If there is an edge $uv$ of $G$ of multiplicity at least three, set its multiplicity to two.
        \item\label{item:fvsfvs_easy_cases2} If there is some $v \in V(G)$ of degree at most one, remove $v$.
        \item\label{item:fvsfvs_easy_cases3} If $u,v \in V(G)$ have $k+2$ common neighbors, add edges $uv$ until $\mul(uv) = 2$.
        \item\label{item:fvsfvs_easy_cases4} If there is some vertex $v$ incident to a loop or to $k+1$ distinct double-edges, remove $v$ and set $k \gets k - 1$.
    \end{enumerate}
\end{restatable}

\begin{sproof}[\cref{rrule:fvsfvs_easy_cases}]
    Let $(G,k)$ be the input instance and $(G', k')$ the instance resulting from the application of this rule.
    We prove each item in order.
    \begin{enumerate}[i.]
        \item Note that at least one of $u,v$ must be in every solution as $\mul(uv) \geq 2$ in both $G$ and $G'$, so $S \in \Sol(G, k)$ if and only if $S \in \Sol(G', k)$.    
        \item Observe that $v$ participates in no cycle of $G$, so $S \in \Sol(G, k)$ if and only if $(S \setminus \{v\}) \in \Sol(G', k)$.
        \item Every solution $S \in \Sol(G, k)$ has one of $u,v$, otherwise we cannot hit all of the at least $k+1$ cycles that contain $u,v$ and two of their neighbors using only $k$ vertices, so it holds that $S \in \Sol(G, k)$ if and only if $S \in \Sol(G',k')$.
        \item Finally, $v$ is mandatory in every solution to $(G,k)$, so $S' \in \Sol(G', k-1)$ if and only if $(S' \cup \{v\}) \in \Sol(G, k)$.
    \end{enumerate}

    In terms of lifting, we only need to be concerned with the second item, as we could have a vertex $v$ in a solution $S \in \Sol(G, k)$ if $S$ is not inclusion-wise minimal. As such, if $S' \in \Sol(G', k)$ has $|S'| < k$, then we must also output $S' \cup \{v\}$, which is in $\Sol(G, k)$.
\end{sproof}

Let us briefly discuss double-edges and their implications to a \pname{Feedback Vertex Set} instance.
Typically, double-edges can be divided into two groups: \emph{real} edges, where we detect that one of the endpoints in fact belongs to every solution, and \emph{virtual} edges, that are used to simplify the instance by showing that at least one solution (if any exists) contains one of the endpoints.
The key reduction rule of Thomassé~\cite{thomasse_fvs}, which yields the required linear degree bound, makes heavy usage of virtual edges; unfortunately, we do not know how to manage the (potentially exponentially) many solutions that are forbidden in the reduced instance by the introduction of these virtual edges, or an alternative rule to linearly bound the maximum degree.
More generally, virtual edges seem to make the lifting process much more complicated.
As such we make a special effort to avoid them.
In particular, this implies that the naive reduction rule that replaces a degree-two vertex in a triangle with a parallel edge between its neighbors is out of the question.
We remark that, while this rule does have an associated lifting procedure, we hope that maintaining only real edges can be leveraged in the future to develop a smaller \pdkernel{} for \pname{Enum Feedback Vertex Set}.

In the following rules, we always assume that we are also given an arbitrary but fixed total ordering $\sigma$ of the vertices of $G$. This is needed to avoid repeatedly outputting a solution during lifting.

\begin{restatable}{fvsfvsrule}{rrulefvsfvsinduceshort}{}
    \label{rrule:fvsfvs_induce_short}
    If three vertices $a,x,b \in V(G)$ form an induced path with $N(x) = \{a,b\}$ and $\sigma(a) < \sigma(b)$, then contract $x$ into $a$.
\end{restatable}

\begin{sproof}[\cref{rrule:fvsfvs_induce_short}]
    For the forward direction, note that every cycle of $G$ containing $x$ also contains both $a$ and $b$.
    For the converse direction, let $(G', k)$ be the resulting instance and $S' \in \Sol(G', k)$.
    Note that $a$ and $b$ play different roles in the rule, and so we must treat them differently in the lifting procedure.
    \begin{enumerate}
        \item\label{item:fvsfvs_induce_short1} If $a \notin S'$, then $S' \in \Sol(G, k)$ and we only output this solution.
        Note that it might have been possible to add $x$ to $S'$; in this case, however, $S' \cup \{a\}$ is a solution of $(G', k)$. To avoid repeated solutions, we defer the choice of adding $x$ to $S'$ to the next case.
        \item\label{item:fvsfvs_induce_short2} If $a \in S'$, then $S' \in \Sol(G, k)$, and again we output it. We have two subcases.
        \begin{enumerate}
            \item\label{item:fvsfvs_induce_short2a} It might have been the case that $S' \cup \{x\} \setminus \{a\}$ is also a solution to $(G,k)$.
            This can be easily checked in polynomial-time by removing $a$ from $S'$ and running any cycle-finding algorithm in $(G' \setminus \{ab\}) \setminus (S' \setminus \{a\})$; if it finds a cycle, then $x$ cannot replace $a$ in $G$, otherwise we output $S' \cup \{x\} \setminus \{a\}$.
            \item\label{item:fvsfvs_induce_short2b} Regardless of the previous outcome, if $|S'| \leq k-1$, then we also output $S' \cup \{x\}$.
        \end{enumerate}
    \end{enumerate}
    It remains to show that the lifting algorithm produces all solutions of $(G,k)$.
    Let $S \in \Sol(G,k)$; we show that it produced by exactly one solution of $G$.
    \begin{itemize}
        \item If $a,x \notin S$, then $S$ is produced in \cref{item:fvsfvs_induce_short1} when given $S$ itself.
        \item If $a \in S$ but $x \notin S$, then $S$ must be output by \cref{item:fvsfvs_induce_short2} when given $S$ itself.
        \item If $a \notin S$ but $x \in S$, then $S' = (S \setminus \{x\}) \cup \{a\}$ is a solution of $(G', k)$. As every cycle containing $x$ contains $a$ and $a \notin S$, it follows that the check in \cref{item:fvsfvs_induce_short2a} succeeds when given $S'$, so it outputs $S$.
        \item Finally, if $a,x \in S$, then $S$ must be output by \cref{item:fvsfvs_induce_short2b} from $S \setminus \{x\}$. \qedhere
    \end{itemize}
\end{sproof}

We note that the following rule does introduce a virtual double-edge, but we remark that, if applied, it will inevitably trigger \cref{rrule:fvsfvs_pending_doubles}, and so we do not violate our own constraints.

\begin{restatable}{fvsfvsrule}{rrulefvsfvsinducedtriangle}{}
    \label{rrule:fvsfvs_induced_triangle}
    If $u, x, y \in V(G)$ form a triangle with $N[x] = N[y] = \{u,x,y\}$, and $\sigma(x) < \sigma(y)$, then contract $y$ into $x$, i.e., make $\mul(ux) = 2$.
\end{restatable}

\begin{sproof}[\cref{rrule:fvsfvs_induced_triangle}]
    The forward direction is trivial, having to at most replace $y$ with $x$ in a solution to the input instance in order to obtain a solution to the compressed instance.
    The converse is also quite direct, as a solution $S'$ of the compressed instance is a solution to the original instance, so we output $S'$. There are, however, two modifications we can make: (\textit{i}) if $x \in S'$, then output $S' \cup \{y\} \setminus \{x\}$; (\textit{ii}) if, additionally, $k - |S'| > 0$, then output $S' \cup \{y\}$. The arguments for the lifting algorithm are very similar to those of \cref{rrule:fvsfvs_induce_short}.
\end{sproof}

\begin{restatable}{fvsfvsrule}{rrulefvsfvspendingdoubles}{}
    \label{rrule:fvsfvs_pending_doubles}
    If $u \in V(G)$ shares double-edges with $t \geq 1$ different vertices $\varepsilon_u = \{v_1, \dots, v_t\}$ with $\deg(v_i) = 2$ for every $i \in [t]$, remove $u$, every $v_i$, and set $k \gets k - 1$.
\end{restatable}

\begin{sproof}[\cref{rrule:fvsfvs_pending_doubles}]
    The forward direction is trivial.
    For the converse, note that we can always add $u$ to the solution $S'$ of the compressed instance $(G', k-1)$ to get a solution to the original instance, and so we output $S' \cup \{u\}$.
    We could, however, have added some of the $v_i$'s, alongside or instead of $u$ to $S'$, as follows:
    \begin{enumerate}
        \item If $t \leq k - |S'|$, then output $S' \cup \varepsilon_u$.
        \item For every non-empty $Z$ subset of size at most $k - |S'| - 1$ of $\varepsilon_u$, output $S' \cup \{u\} \cup Z$.
    \end{enumerate}
    Note that the second case may be applied even if the \verb|if| condition of the first one is satisfied.
    Furthermore, the non-empty requirement in the second case guarantees that $S' \cup \{u\}$ is not output by it.
\end{sproof}

\begin{restatable}{fvsfvsrule}{rrulefvsfvsmultiflags}{}
    \label{rrule:fvsfvs_multiflags}
    If there is a pair of vertices $a,b \in V(G)$ and a set of degree-two vertices $\varepsilon_{ab} = \{x_1,\dots, x_t\}$ with $N(x_i) = \{a,b\}$ for every $i \in [t]$ and $N(b) = \{a\} \cup \varepsilon_{ab}$, then remove $N[b]$ and set $k \gets k - 1$.
\end{restatable}

\begin{sproof}[\cref{rrule:fvsfvs_multiflags}]
    The forward direction is immediate.
    For the converse, let $S'$ be a solution to the compressed instance $(G', k-1)$.
    As in the proof of \cref{rrule:fvsfvs_pending_doubles} we have some cases:
    \begin{itemize}
        \item We output $S' \cup \{a\}$; it is a solution to $(G, k)$ as every cycle in $G$ not present in $G'$ contains $a$ and $|S'| \leq k - 1$.
        \item For each non-empty subset $Z$ of $\varepsilon_{ab} \cup \{b\}$ of size $k - |S'| - 1$, we output $S' \cup \{a\} \cup Z$.
        \item If $(G \setminus S') \setminus (N[b] \setminus \{a\})$ is acyclic, then the only cycles introduced by adding $N[b]$ back into $G'$ are precisely the triangles containing $a$, $b$, and one vertex of $\varepsilon_{ab}$.
        As such, we have two subcases:
        \begin{itemize}
            \item For each subset $Z$ of $\varepsilon_{ab}$ of size at most $k - |S'| - 1$, output $S' \cup \{b\} \cup Z$.
            \item If $|\varepsilon_{ab}| \leq k - |S'|$, output $S' \cup \varepsilon_{ab}$.
        \end{itemize}
    \end{itemize}
    Note that, when computing the subsets $Z$ in the above case analysis, the textbook branching algorithm that at the $i$-th branching step either adds $x_i$ or forbids $x_i$ is a linear-delay algorithm.
\end{sproof}

\begin{restatable}{lemma}{lemfvsfvsdeglowerbound}{}
    \label{lem:fvsfvs_deg_lower_bound}
    Let $G$ be a graph and $\Delta$ its maximum degree.
    If none of the \cref{rrule:fvsfvs_easy_cases,rrule:fvsfvs_induce_short,rrule:fvsfvs_induced_triangle,rrule:fvsfvs_pending_doubles,rrule:fvsfvs_multiflags} are applicable, then $\fvs(G) \geq |V(G)|/(3\Delta-3)$.
\end{restatable}

\begin{proof}
    By \cref{rrule:fvsfvs_easy_cases}, no vertex of $G$ has degree zero or one.
    Let $V_2$ be the set of degree-two vertices of $G$, and $V_{\geq 3} = V(G) \setminus V_2$.

    \begin{claim}
        \label{clm:fvsfvs_deg_lower_bound_edgeless}
        The graph $G[V_2]$ is edgeless.
    \end{claim}
    
    \begin{cproof}
        Towards a contradiction, take $u,v \in V_2$ and suppose $uv \in E(G[V_2])$.
        As \cref{rrule:fvsfvs_pending_doubles} is not applicable, $\mul(uv) = 1$.
        If, on the other hand, there is some $z \in V(G)$ with  $N[u] = N[v] = \{u,v,z\}$, then \cref{rrule:fvsfvs_induced_triangle} could be applied.
        Finally, if $u$ and $v$ have no common neighbor, then we would have applied \cref{rrule:fvsfvs_induce_short}.
    \end{cproof}

    \begin{claim}
        \label{clm:fvsfvs_deg_lower_bound_deg2}
        Every vertex of $G[V_{\geq 3}]$ has degree at least two in $G[V_{\geq 3}]$.
    \end{claim}

    \begin{cproof}
        We first show that there are no isolated vertices in $G[V_{\geq 3}]$.
        Assume that $v$ is isolated and note that, since $\deg(v) \geq 3$, the set $U =\{u_1, \dots, u_t\}$ of degree-two neighbors of $v$ in $G$ has at least three vertices, $N(v) = U$, and $G[U]$ is edgeless since $G[V_2]$ is independent and $U \subseteq V_2$.
        Moreover, $\mul(u_iv) = 1$ for every $u_i \in U$ as \cref{rrule:fvsfvs_pending_doubles} is not applicable.
        As $U$ is an independent set, there is at least one other vertex $z \in N_G(U) \setminus \{v\}$.
        As \cref{rrule:fvsfvs_induce_short} is not applicable, $vz \in E(G)$; this, however, either: (a) contradicts that $G[U]$ is an independent set if $\deg_G(z) = 2$, or (b) that $N_G(v) = U$, so $v$ must not have been isolated in $G[V_{\geq 3}]$.

        Let $v \in V_{\geq 3}$, $U = N_G(v) \cap V_2$, and $z \in V_{\geq 3} \cap N_{G}(v)$; by our previous paragraph, $z$ exists.
        If $U \subseteq N_G(z)$ or $N_G(z) \subseteq U \cup \{v\}$, then \cref{rrule:fvsfvs_multiflags} is applicable to $v,z,U$.
        Thus, there is some $u \in U \setminus N_G(z)$ with a neighbor $y \in V_{\geq 3} \setminus \{v, z\}$.
        If $vy \notin E(G[V_{\geq 3}])$, then \cref{rrule:fvsfvs_induce_short} would be applicable; we present an example of this situation in \cref{fig:fvs_v2}.
        Finally, we conclude that  $vy \in E(G[V_{\geq 3}])$ and it follows that $|N(v) \cap V_{\geq 3}| \geq 2$, contradicting that $v$ has only one neighbor in $V_{\geq 3}$. \qedhere
        
        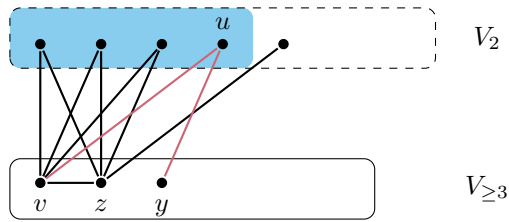
\begin{figure}[!htb]
            \centering
            \begin{tikzpicture}[scale=0.8]
              \GraphInit[unit=3,vstyle=Normal]
              \SetVertexNormal[Shape=circle, FillColor=black, MinSize=2pt]
              \tikzset{VertexStyle/.append style = {inner sep = \inners, outer sep = \outers}}
              \SetVertexLabelOut
            
              \begin{scope}[yshift=3cm]
                \draw[dashed, rounded corners] (-3, -0.5) rectangle (4, 0.5);
                
                \node at (4.85, 0) {$V_2$};
                  \foreach \i in {1,...,5} {
                    \pgfmathsetmacro{\x}{\i-3.5}
                    \Vertex[x=\x, y=-0.1,NoLabel]{c\i}
                  }
                  \node at ($(c4) + (0,0.35)$) {$u$};
                  \begin{scope}[on background layer]
                      \fill[goodcyan, rounded corners] ($(c1)-(0.5, 0.4)$) rectangle ($(c4)+(0.5, 0.6)$);
                  \end{scope}
              \end{scope}
              
              \begin{scope}
                \draw[rounded corners] (-3, 0) rectangle (3, 1);
                \node at (4.85, 0.5) {$V_{\geq 3}$};
                \Vertex[x=-2.5, y=0.6, LabelOut, Lpos=-90, Math]{v}
                \Vertex[x=-1.5, y=0.6, LabelOut, Lpos=-90, Math]{z}
                \Vertex[x=-0.5, y=0.6, LabelOut, Lpos=-90, Math]{y}
                \Edges(c5,z,c1,v,c2,z,c3,v,z)
                \SetUpEdge[color=goodred]
                \Edges(y,c4,v)
              \end{scope}
            \end{tikzpicture}
            \caption{Case contradicting the inapplicability of \cref{rrule:fvsfvs_induce_short} in the proof of \cref{lem:fvsfvs_bounding_lemma}. The four cyan-shaded vertices correspond to $U = N(v) \cap V_2$, while the two red edges witness the applicability of \cref{rrule:fvsfvs_induce_short} if $vy \notin E(G[V_{\geq 3}])$.\label{fig:fvs_v2}}
        \end{figure}

    \end{cproof}
    
    Now, let $X$ be a minimum feedback vertex set of $G$, $F = G \setminus X$, and $E_X$ the set of edges incident to at least one vertex of $X$.
    Note that, $E(G) = E_X \cup E(G[F])$, which implies that:
    
    \begin{equation}
        \label{eq:fvs_m_n}
        \Delta\cdot\fvs(G) \geq |E_X| \geq |E(G)| - (|V(G)| - \fvs(G) - 1).
    \end{equation}

    Let us now lower bound $|E(G)|$ by a function of $|V(G)|$ and $\Delta$; this is accomplished by counting the edges of the graph using a scheme inspired by the accounting method of amortized analysis and the discharging method~\cite{cranston_discharging}, commonly used in graph coloring problems.
    To do so, we must set a weight $w: V(G) \mapsto \mathbb{Q}$ to each vertex in such a way that an edge $uv \in E(G)$ charges $u$ with a fraction $1/c_{uv}$ and charges $v$ with $1 - 1/c_{uv}$, where $c_{uv} \geq 1$ is a rational number. This avoids over-counting $uv$ when summing the edges; $w(v)$ is defined as the sum of the charges given to $v$.
    Typically, one would set $c_{uv} = 2$ and have $|E(G)| = \sum_{v \in V(G)}\deg(v)/2$; we will simply move the weights around to get our lower bound.
    We do this as follows:
    \begin{itemize}
        \item If $v \in V_{\geq 3}$, then each edge $uv$ incident to $v$ in $G[V_{\geq 3}]$ gets $c_{uv} = 2$ and contributes with $1/2$ to both $w(u)$ and $w(v)$.
        \item By \cref{clm:fvsfvs_deg_lower_bound_edgeless}, $G[V_2]$ is edgeless; consequently, all edges incident to $v \in V_2$ have the other endpoint in $V_{\geq 3}$.
        As such, for each $uv \in E(G)$, we set $c_{uv} = 3$, charge $w(u)$ with $1/3$ and $w(v)$ with $2/3$,
    \end{itemize}
    Now, let $A = \{v \in V_{\geq 3} \mid |N(v) \cap V_{\geq 3}| = 2\}$, $B = V_{\geq 3} \setminus A$, $n_A = |A|$, $n_B = |B|$, $n_2 = |V_2|$, and $\delta_B = \min\{|N(v) \cap V_{\geq 3}| \mid v \in B\}$; if $B = \emptyset$, $\delta_B = 0$. By \cref{clm:fvsfvs_deg_lower_bound_deg2} and the definition of $B$, $\delta_B \geq 3$.
    Let us analyze how the weights behave.
    \begin{itemize}
        \item If $v \in B$, then $w(v) \geq \delta_B/2 \geq 3/2 \geq 4/3$, as each edge incident to $v$ in $G[V_{\geq 3}]$, of which there are at least $\delta_B$, contributes with $1/2$ units of weight to $w(v)$.
        \item If $v \in A$, then, by definition, it is incident to exactly two edges in $G[V_{\geq 3}]$; since $\deg_G(v) \geq 3$, it is adjacent to at least one vertex in $V_2$, so we have $w(v) \geq 2\cdot 1/2 + 1/3 = 4/3$.
        \item Finally, if $v \in V_2$, then it was charged with $2/3$ units of weight for each edge incident to it; since $G[V_2]$ is edgeless, no edge is overcounted, and so $w(v) = 4/3$.
    \end{itemize}
    Thus, we have that:
    \begin{equation*}
        |E(G)| \geq \frac{4}{3} (n_A + n_B + n_2)
          \geq \frac{4}{3}n.
    \end{equation*}
    The above, along with \cref{eq:fvs_m_n}, allows us to conclude that:

    \begin{align*}
        \Delta\cdot \fvs(G) &\geq \frac{4}{3}n - n + \fvs(G)\\ 
        \fvs(G) &\geq \frac{n}{3(\Delta -1)}. \qedhere
    \end{align*}
\end{proof}

With \cref{lem:fvsfvs_deg_lower_bound} at hand, we are  tasked with bounding the maximum degree of $G$.
Formally, a \emph{$v$-flower of order $t$} is a family of $t$ cycles that pairwise intersect only at vertex $v$.
\cref{rrule:fvsfvs_flower} and \cref{lem:flower_finder} are taken straight from~\cite{thomasse_fvs}; we omit the proof of the latter for brevity, while we give a proof for the former as we must discuss how to lift the solutions of the reduced instance back to the input instance.

\begin{restatable}{fvsfvsrule}{rrulefvsfvsflower}{}
    \label{rrule:fvsfvs_flower}
    If there is a vertex $v \in V(G)$ and an $v$-flower of order at least $k+1$, remove $v$ and set $k \gets k - 1$.
\end{restatable}

\begin{sproof}[\cref{rrule:fvsfvs_flower}]
    The correctness follows immediately from the fact that $v$ is a mandatory vertex in every feedback vertex set of size at most $k$ of $G$.
    As such, lifting from a solution $X'$ of the reduced instance amounts to outputting $X' \cup \{v\}$.
\end{sproof}

\begin{lemma}[\!\!\cite{thomasse_fvs}]
    \label{lem:flower_finder}
    Let $G$ a multigraph, $X$ a feedback vertex set of $G$ of size at most $2k$, and $x \in V(G)$. In polynomial time, it is possible to accomplish exactly one of the following:
    \begin{enumerate}
        \item Decide if $(G,k)$ has $\Sol(G, k) = \emptyset$, i.e., the corresponding decision problem is a \NOi-instance;
        \item Find a $v$-flower of order $k+1$; or
        \item Find a set feedback vertex set $H_x$ such that $X \setminus \{x\} \subseteq H_x \subseteq V(G) \setminus \{x\}$ of size at most $3k$.
    \end{enumerate}
\end{lemma}

What is left now is to show how to leverage the third property of \cref{lem:flower_finder}.
To do so, suppose that \cref{rrule:fvsfvs_flower} is not applicable, take $v \in V(G)$ with $\deg(v) > 3k(k+1) + 5k$, let $H_v$ be as in \cref{lem:flower_finder}, and let $\C$ denote the set of connected components of $G \setminus (H_v \cup \{v\})$.
Note that, as $X \subseteq H_v \cup \{v\}$, each $C_i \in \C$ is a tree.

\begin{observation}
    \label{obs:fvsfvs_auxiliary_structure}
    There exists $\D_v \subseteq \C$ with $|\D_v| > 3k(k+1)$ such that, for every $D_i \in \D_v$: $v$ is adjacent to one vertex of $D_i$ with a non-double-edge, and $D_i$ has at least one vertex adjacent to a vertex of $H_v$.
\end{observation}

\begin{proof}
    As $H_v$ hits all cycles containing $v$, $v$ has at most one neighbor in each $C_i \in \C$; note that $H_v$ contains every vertex incident to a double-edge incident to $v$.
    Due to the fact that $|H_v| \leq 3k$ and that \cref{rrule:fvsfvs_easy_cases} is not applicable, $H_v$ contributes with at most $5k$ to $\deg(v)$, implying that more than $3k(k+1)$ elements of $\C$ contain a vertex in $N(v)$ as $\deg(v) > 3k(k+1) + 5k$; let $\D_v = \{D_1, \dots, D_t\}$ be this collection.
    Take any $D_i \in \D_v$.
    For the last statement, we have two cases.
    First, if $|D_i| = 1$, then the inapplicability of \cref{rrule:fvsfvs_easy_cases} guarantees that $u \in D_i$ has at least one more neighbor, which must be in $H_v$.
    Second, if $|D_i| \geq 2$, then $D_i$ has at least two leaves $u_1, u_2$ and at most one of them is adjacent to $v$, otherwise there would be a cycle containing $v$ and avoiding $H_v$.
    Consequently, we have that the other leaf is adjacent to a vertex of $H_v$ or \cref{rrule:fvsfvs_easy_cases} would be applicable.
\end{proof}

At this point, we essentially have the same setup as Thomassé~\cite{thomasse_fvs}.
Instead of leveraging $q$-expansions, however, we take a simpler approach, which does not yield a linear bound but allows us to obtain a polynomial \pdkernel{}; we remark that it is still interesting to explore if these generalizations of crown decompositions can be leveraged to reduce the quadratic bound on the maximum degree that we obtain.
We are first going to slightly increase the degree of $v$, before decreasing it, as follows.

\begin{restatable}{fvsfvsrule}{rrulefvsfvsauxiliarydouble}{}
    \label{rrule:fvsfvs_auxiliary_double}
    Let $\D_v, H_v$ be as above, and $G_v$ be the bipartite graph with bipartition $(\{d_i \mid D_i \in \D_v\}, \{h_u \mid u \in H_v\})$, and $d_ih_u \in E(G_v)$ if and only if $N_G(u) \cap D_i \neq \emptyset$.
    For every $h_u \in V(G_v)$ of degree at least $k+2$, add a double-edge $uv$ to $G$.
\end{restatable}

\begin{sproof}[\cref{rrule:fvsfvs_auxiliary_double}]
    The proof follows immediately from the observation that, if neither $u$ nor $v$ are picked in a solution, then for all but one $D_i$ such that $d_i \in N_{G_v}(h_u)$, we must pick a vertex of $D_i$ in a solution to $(G,k)$, otherwise there is a cycle containing $u$, $v$, and two components $D_i$, $D_j$ with $d_i,d_j \in N_{G_v}(h_u)$.
    As this would imply picking $k+1$ vertices, we obtain that every solution contains at least one element of $\{v, u\}$.
\end{sproof}

We remark that the double-edges introduced by \cref{rrule:fvsfvs_auxiliary_double} are not virtual: there are $k+2$ cycles with pairwise intersection equal to $\{u,v\}$.

\begin{restatable}{fvsfvsrule}{rrulefvsfvsedgedeletion}{}
    \label{rrule:fvsfvs_edge_deletion}
    Let $H^2_v$ be the vertices of $H_v$ with a double-edge to $v$.
    If there exists $D_i \in \D_v$ with $N_G(D_i) \setminus \{v\} \subseteq H^2_v$, remove $vw$ from $G$, where $w \in D_i \cap N_G(v)$.
\end{restatable}

\begin{sproof}[\cref{rrule:fvsfvs_edge_deletion}]
    Let $(G', k)$ be the output instance.
    For the forward direction observe that any solution to $(G,k)$ is a solution to $(G', k)$, as we only destroy cycles by removing edges.
    For the converse direction, take $S' \in \Sol(G', k)$.
    We claim that $S' \in \Sol(G,k)$; indeed, every cycle in $G$ that does not exist in $G'$ contains $v,w$ and some vertex of $H^2_v$ as $N_G(D_i) \setminus \{v\} \subseteq H^2_v$, but we know that either $v \in S'$ or the other endpoints of all of its double-edges are in $S'$, so such a cycle cannot exist in $G \setminus S'$. \qedhere
\end{sproof}

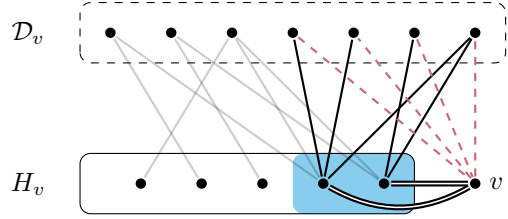
\begin{figure}[!htb]
    \centering
    \begin{tikzpicture}[scale=0.8]
      \GraphInit[unit=3,vstyle=Normal]
      \SetVertexNormal[Shape=circle, FillColor=black, MinSize=2pt]
      \tikzset{VertexStyle/.append style = {inner sep = \inners, outer sep = \outers}}
      \SetVertexNoLabel
    
      \begin{scope}[yshift=3cm]
        \draw[dashed, rounded corners] (-3, -0.5) rectangle (4, 0.5);
        
        \node at (-3.85, 0) {$\D_v$};
          \foreach \i in {1,...,7} {
            \pgfmathsetmacro{\x}{\i-3.5}
            \Vertex[x=\x, y=0,NoLabel]{c\i}
          }
      \end{scope}
      
      \begin{scope}
        \draw[rounded corners] (-3, 0) rectangle (2.5, 1);
        \node at (-3.85, 0.5) {$H_v$};
        \foreach \i in {1,...,5} {
          \pgfmathsetmacro{\x}{\i-3}
          \Vertex[x=\x, y=0.5,NoLabel]{h\i}
        }
        \Vertex[x=3.5, y=0.5,NoLabel]{v}
        \node at ($(v) + (0.35,0)$) {$v$};
        \Edges[style={double}](v,h5)
        \Edges[style={bend left, double}](v,h4)
        \begin{scope}[on background layer]
            \fill[goodcyan, rounded corners] ($(h4)-(0.5, 0.5)$) rectangle ($(h5)+(0.5, 0.5)$);
        \end{scope}

          \Edges[style={opacity=0.2}](h2,c1,h4,c3)
          \Edges[style={opacity=0.2}](h3,c2,h5,c3,h1)
          \Edges(c4,h4,c7,h5,c6)
          \Edges(c5,h4)
          \SetUpEdge[color=goodred,style={dashed}]
          \Edges(c4,v,c5)
          \Edges(c6,v,c7)
      \end{scope}
    \end{tikzpicture}
    \caption{Example of repeated applications of \cref{rrule:fvsfvs_edge_deletion}; where each $D_i \in \D_v$ is contracted into a single vertex for ease of presentation. The two cyan-shaded vertices correspond to $H^2_v$. The double-edges incident to $v$ are real double-edges, the seven light-gray edges are incident to vertices of $\D_v$ not affected by the rule, the four red dashed edges are the ones removed by the reduction rule, and the five solid edges are incident to vertices of $\D_v$ affected by the rule.\label{fig:fvs_edge_deletion}}
\end{figure}

\cref{rrule:fvsfvs_edge_deletion} is our key rule to reduce the maximum degree of the graph; we present an example of repeated applications of it in \cref{fig:fvs_edge_deletion}.
We are now ready to bound the maximum degree of $G$ by a function of $k$; along with \cref{lem:fvsfvs_deg_lower_bound}, this will imply a bound on the overall size of $G$.

\begin{lemma}
    \label{lem:fvsfvs_bounding_lemma}
    If \cref{rrule:fvsfvs_auxiliary_double,rrule:fvsfvs_edge_deletion} are not applicable, then $G$ has maximum degree at most $3k(k+1) + 5k$.
\end{lemma}

\begin{proof}
    Suppose that there is some $v \in V(G)$ incident to more than $3k(k+1) + 5k$ edges.
    By \cref{obs:fvsfvs_auxiliary_structure} and \cref{rrule:fvsfvs_flower}, the set $\D_v$ of connected components of $G \setminus (H_v \cup \{v\})$ that $v$ has a neighbor in, but no double-edges to, has size greater than $3k(k+1)$; moreover, every $D_i \in \D_v$ has a neighbor in $H_v$.
    Now, take some $u \in H_v$ for which $\mul(uv) < 2$; as \cref{rrule:fvsfvs_auxiliary_double} is not applicable, we know that $N_G(u)$ hits at most $k+1$ components of $\D_v$.
    As such, taking $H^2_v$ as in \cref{rrule:fvsfvs_edge_deletion}, we know that $N_G(H_v \setminus H^2_v)$ hits at most $|H_v \setminus H^2_v|\cdot (k+1)$ components of $\D_v$.
    Note that, if $H^2_v = \emptyset$ and $|\D_v| > |H_v|\cdot (k + 1)$, then there is some $h \in H_v$ adjacent to vertices in at least $k+2$ components of $\D_v$ and \cref{rrule:fvsfvs_auxiliary_double} is applicable, so it must be the case that $H^2_v \neq \emptyset$.
    As such, since at most $(|H_v| - 1)(k+1)$ components of $\D_v$ have a vertex adjacent to a vertex in $H_v \setminus H^2_v$, it follows that there is at least one component of $\D_v$ whose neighborhood is contained in $H^2_v \cup \{v\}$, which is enough to trigger \cref{rrule:fvsfvs_edge_deletion}.
\end{proof}

\begin{lemma}
    \label{lem:fvsfvs_compression}
    There is a polynomial-time algorithm that, given an instance $(G,k)$ of \pname{Enum Feedback Vertex Set}, either answers that $\Sol(G, k) = \emptyset$ or outputs an equivalent instance $(G', k')$ where $|V(G)| \leq 3k^3 + 8k^2$.
\end{lemma}

\begin{proof}
    The algorithm simply consists of the exhaustive application of \cref{rrule:fvsfvs_easy_cases,rrule:fvsfvs_induce_short,rrule:fvsfvs_induced_triangle,rrule:fvsfvs_pending_doubles,rrule:fvsfvs_multiflags,rrule:fvsfvs_flower,rrule:fvsfvs_auxiliary_double,rrule:fvsfvs_edge_deletion}, which also guarantee that $(G,k)$ and $(G', k')$ are equivalent.
    By \cref{lem:fvsfvs_bounding_lemma}, we have that the maximum degree $\Delta$ of $G'$ is at most $3k(k+1) + 5k$.
    By \cref{lem:fvsfvs_deg_lower_bound}, we have that $\fvs(G') \geq |V(G')|/(3\Delta - 3)$.
    As such, if $k'(3k'(k'+1) + 5k') < |V(G')|$, then we know that no feedback vertex set of size $k'$ exists in $G'$, and so $\Sol(G', k') = \Sol(G, k) = \emptyset$.
    Otherwise, $|V(G')| \leq 3k'^3+3k'^2 + 5k'^2 = 3k'^3 + 8k'^2$ and we are done.
\end{proof}

\begin{lemma}
    \label{lem:fvsfvs_lifting}
    There is a polynomial-delay algorithm that, given $(G, k)$, $(G', k')$, and $S' \in \Sol(G', k')$, returns a non-empty set $\Lift(S') \subseteq \Sol(G, k)$.
    Moreover, $\Lift(G', k) \coloneq \{\Lift(S') \mid S' \in \Sol(G', k')\}$ is a partition of $\Sol(G, k)$.
\end{lemma}

\begin{proof}
    The lifting algorithm begins by running the compression algorithm of \cref{lem:fvsfvs_compression} on $(G, k)$; as no step is randomized, we have the exact same applications of the reduction rules used to obtain $(G', k')$.
    As to how to perform the lifting of $S'$, we use the methods described in the converse part of each of our reduction rules but, instead of enumerating all possibilities at once, when a viable solution is found, we immediately proceed with the recursion.
    This leads to a polynomial-delay algorithm as every reduction rule has a polynomial-delay lifting algorithm and the recursion tree described above has polynomial depth.
    As each of our rules either keeps $S'$ as it or only adds vertices to $S'$ that do not belong to $G'$, it immediately follows that our kernel is extension-only, so no other $S''\in \Sol(G', k')$ has $\Lift(S'') \cap \Lift(S') \neq \emptyset$, otherwise we would conclude that $S' = S''$.
    Finally, to observe that $\Lift(G', k)$ is indeed a partition of $\Sol(G,k)$, note that, by the forward direction of our reduction rules, every $S \in \Sol(G, k)$ is assigned to a unique solution of $(G', k')$, and this solution is obtained simply by removing vertices from $S$.
\end{proof}

Finally, the proof of \cref{thm:fvsfvs_strong_kernel} follows immediately from \cref{lem:fvsfvs_compression,lem:fvsfvs_lifting}.

\fvsfvsstrongkernel*
\section{Final remarks}
\label{sec:final}

In this paper, we have developed polynomial-sized  \pdkernels{} for the natural parameterizations of enumeration variants of some key parameterized problems, namely \pname{Enum Vertex Cover} and \pname{Enum Feedback Vertex Set}.
In this process, we showed how to leverage crown decompositions, an important technique initially developed for decision kernelization, to the enumeration setting.
We hope that these results can further motivate the study of this new domain of parameterized complexity.
Among concrete open problems, we are interested in the enumeration of minimal vertex covers and feedback vertex sets; we believe our techniques are applicable to these variants of the studied problems.
A more challenging question is the existence of a quadratic kernel for \pname{Enum Feedback Vertex Set}, which we do not see how to obtain using our approach; it seems like tools describing global structures of the graph, like $q$-expansions, are needed.
Finally, we remark that lower bound theories for enumerative kernelization and enumerative parameterized complexity would be of extreme benefit to the area.
Currently, lower bounds are derived almost solely via the related parameterized {\sl decision} problems.
It is unlikely, however, that this strategy is enough to (conditionally) classify parameterized enumeration problems into tractable and intractable and which admit polynomial \pdkernels{} and those that do not.

\bibliography{refs}

\end{document}